\documentclass[twoside,twocolumn,9pt]{article}
\usepackage{extsizes}
\usepackage[super,sort&compress,comma]{natbib} 
\usepackage[version=3]{mhchem}
\usepackage[left=1.5cm, right=1.5cm, top=1.785cm, bottom=2.0cm]{geometry}
\usepackage{balance}
\usepackage{mathptmx}
\usepackage{sectsty}
\usepackage{graphicx} 
\usepackage{lastpage}
\usepackage[format=plain,justification=justified,singlelinecheck=false,font={stretch=1.125,small,sf},labelfont=bf,labelsep=space]{caption}
\usepackage{float}
\usepackage{fancyhdr}
\usepackage{fnpos}
\usepackage[english]{babel}
\addto{\captionsenglish}{%

}
\usepackage{array}
\usepackage{droidsans}
\usepackage{charter}
\usepackage[T1]{fontenc}
\usepackage[usenames,dvipsnames]{xcolor}
\usepackage{setspace}
\usepackage[compact]{titlesec}
\usepackage{hyperref}

\usepackage{epstopdf}

\definecolor{cream}{RGB}{222,217,201}

\usepackage{bm}
\usepackage{comment}

\begin{document}

\pagestyle{fancy}
\thispagestyle{plain}
\fancypagestyle{plain}{
\renewcommand{\headrulewidth}{0pt}
}

\newcommand*{\usnote}[1]{{\color{red}({\bf \footnotesize US: } #1)}}
\newcommand*{\fznote}[1]{{\color{blue}({\bf \footnotesize FZ: } #1)}}
\newcommand*{\rcnote}[1]{{\color{olive}({\bf \footnotesize RC: } #1)}}
\newcommand{\matr}[1]{\boldsymbol{\bm{#1}}}
\renewcommand{\vec}{\bm}
\newcommand*{\rev}[1]{{\color{red}#1}}

\makeFNbottom
\makeatletter
\renewcommand\LARGE{\@setfontsize\LARGE{15pt}{17}}
\renewcommand\Large{\@setfontsize\Large{12pt}{14}}
\renewcommand\large{\@setfontsize\large{10pt}{12}}
\renewcommand\footnotesize{\@setfontsize\footnotesize{7pt}{10}}
\makeatother

\renewcommand{\thefootnote}{\fnsymbol{footnote}}
\renewcommand\footnoterule{\vspace*{1pt}%
\color{cream}\hrule width 3.5in height 0.4pt \color{black}\vspace*{5pt}} 
\setcounter{secnumdepth}{5}

\makeatletter 
\renewcommand\@biblabel[1]{#1}            
\renewcommand\@makefntext[1]%
{\noindent\makebox[0pt][r]{\@thefnmark\,}#1}
\makeatother 
\renewcommand{\figurename}{\small{Fig.}~}
\sectionfont{\sffamily\Large}
\subsectionfont{\normalsize}
\subsubsectionfont{\bf}
\setstretch{1.125} 
\setlength{\skip\footins}{0.8cm}
\setlength{\footnotesep}{0.25cm}
\setlength{\jot}{10pt}
\titlespacing*{\section}{0pt}{4pt}{4pt}
\titlespacing*{\subsection}{0pt}{15pt}{1pt}

\fancyfoot{}
\fancyfoot[LO,RE]{\vspace{-7.1pt}\includegraphics[height=9pt]{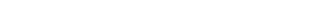}}
\fancyfoot[CO]{\vspace{-7.1pt}\hspace{13.2cm}\includegraphics{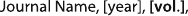}}
\fancyfoot[CE]{\vspace{-7.2pt}\hspace{-14.2cm}\includegraphics{head_foot/RF}}
\fancyfoot[RO]{\footnotesize{\sffamily{1--\pageref{LastPage} ~\textbar  \hspace{2pt}\thepage}}}
\fancyfoot[LE]{\footnotesize{\sffamily{\thepage~\textbar\hspace{3.45cm} 1--\pageref{LastPage}}}}
\fancyhead{}
\renewcommand{\headrulewidth}{0pt} 
\renewcommand{\footrulewidth}{0pt}
\setlength{\arrayrulewidth}{1pt}
\setlength{\columnsep}{6.5mm}
\setlength\bibsep{1pt}

\makeatletter 
\newlength{\figrulesep} 
\setlength{\figrulesep}{0.5\textfloatsep} 

\newcommand{\topfigrule}{\vspace*{-1pt}%
\noindent{\color{cream}\rule[-\figrulesep]{\columnwidth}{1.5pt}} }

\newcommand{\botfigrule}{\vspace*{-2pt}%
\noindent{\color{cream}\rule[\figrulesep]{\columnwidth}{1.5pt}} }

\newcommand{\dblfigrule}{\vspace*{-1pt}%
\noindent{\color{cream}\rule[-\figrulesep]{\textwidth}{1.5pt}} }

\makeatother

\twocolumn[
  \begin{@twocolumnfalse}
{\includegraphics[height=30pt]{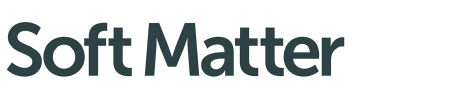}\hfill\raisebox{0pt}[0pt][0pt]{\includegraphics[height=55pt]{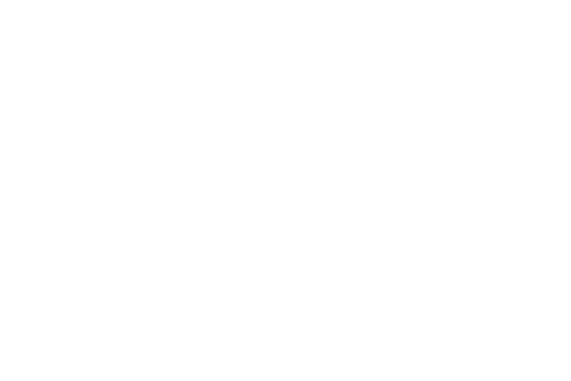}}\\[1ex]
\includegraphics[width=18.5cm]{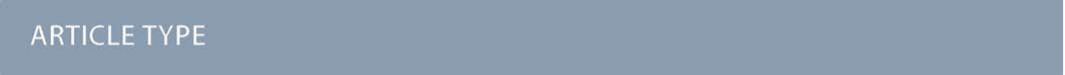}}\par
\vspace{1em}
\sffamily
\begin{tabular}{m{4.5cm} p{13.5cm} }

\includegraphics{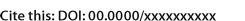} & \noindent\LARGE{\textbf{The role of the nucleus for cell mechanics: an elastic phase field approach}} \\
\vspace{0.3cm} & \vspace{0.3cm} \\

 & \noindent\large{Robert Chojowski,\textit{$^{a,b}$} Ulrich S. Schwarz\textit{$^{a,b}$} and Falko Ziebert$^{\ast}$\textit{$^{a,b}$}} 
 \\

\includegraphics{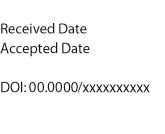} & \noindent\normalsize{The 
nucleus of eukaryotic cells typically makes up around 30\% of the cell volume and has significantly different mechanics, which can make it effectively 
up to ten times stiffer than the surrounding cytoplasm.
Therefore it is an important element for cell mechanics, but a quantitative
understanding of its mechanical role during whole cell dynamics is largely missing. 
Here we demonstrate that elastic phase fields can be used to describe dynamical 
cell processes in adhesive or confining environments in which the
nucleus acts as a stiff inclusion. 
We first introduce and verify our computational method and then study several applications of large relevance. 
For cells on adhesive patterns, we find that nuclear stress is shielded
by the adhesive pattern. 
For cell compression between two parallel plates, we obtain force-compression curves that allow us to extract an effective modulus for the cell-nucleus composite. 
For micropipette aspiration, the effect of the nucleus on the effective modulus is found to be much weaker, highlighting the complicated interplay between extracellular geometry and cell mechanics that is captured by our approach.
We also show that our phase field approach can be used to investigate
the effects of viscoelasticity and cortical tension.
} \\
\end{tabular}

 \end{@twocolumnfalse} \vspace{0.6cm}

  ]

\renewcommand*\rmdefault{bch}\normalfont\upshape
\rmfamily
\section*{}
\vspace{-1cm}


\footnotetext{\textit{$^{a}$~Institute for Theoretical Physics, Heidelberg University, Philosophenweg 19, 69120 Heidelberg, Germany; E-mail: f.ziebert@thphys.uni-heidelberg.de}}

\footnotetext{\textit{$^{b}$~BioQuant, Heidelberg University, Im Neuenheimer Feld 267, 69120 Heidelberg, Germany}}




\section{Introduction}

Many essential biological processes depend on the mechanical properties of animal cells
and their ability to dynamically react to mechanical cues from their environment. 
Classical examples include the spreading behaviour of cells on substrates of variable stiffness,~\cite{PelhamWang,nisenholz_active_2014} 
cell migration in the direction of larger stiffness~\cite{duroDemboWang,sunyer_collective_2016} and cell differentiation
in response to environmental stiffness.~\cite{mcbeath_cell_2004,Engler_2006_Matrix_stiffness_guides_cell_differentiation}
A typical cell response to variable environmental stiffness is to adapt the own stiffness to match the
one of the environment.~\cite{solon_fibroblast_2007,schwarz_soft_2007} 
However, there are also situations in which it is favorable for cells to work with a different stiffness then the surrounding. 
One prominent example are migratory immune and cancer cells in confined spaces,
which tend to increase their softness in order to more easily squeeze through the pores in their environment. \cite{Guck_2005_optical_deformability_as_inherent_Cell_marker_for_testing_malignant_transformation_and_metastatic_competence,Thiam_2016_Perinuc_Arp2_3_actin_polym_enables_nucl_deformation_to_faciitates_cell_migration,fuhs_rigid_2022}

The main determinant of cell mechanics is the cytoskeleton, a 
crosslinked and highly dynamical polymer network, 
giving the cell stability and the ability to quickly change 
its mechanics.~\cite{Stricker_2010_review_actin_CSK,Murrell_2015_review_actinCSK,Pegoraro_2017_review_mechanics_CSK} 
In particular, the cytoskeleton allows cells to 
generate forces, mainly pushing forces through polymerization and
pulling forces through motor activity, both of which convert chemical
energy into mechanical work and thus make the cell an active system.~\cite{Schwarz_Safran_2013,Murrell_2015_review_actinCSK} 
Although the plasma membrane typically does not contribute much 
to cell mechanics directly, it is important in the sense that it determines
cell volume and surface area; in addition, it provides guidance for
the organization of the cell cortex generated by the cytoskeleton
as a thin polymeric network wrapping the whole cell.~\cite{salbreux_actin_2012,Svitkina_2020_cortex_review}

In recent years, it has become clear that a third important mechanical component
of animal cells is the nucleus. \cite{Kalukula_2022_mechanics_and_functional_consequences_of_nuclear_deformations}
The nucleus harbours the genetic information of the cell and is separated from the cytoplasm by its nuclear envelope.
Due to its overarching role for gene expression, it has long been overlooked that the nucleus also plays an important role in mechanics. 
Having a cell-type dependent diameter of several micrometers and occupying a large fraction of the overall cell volume (typically 
up to 30\%), the nucleus is the largest and most prominent of all cellular organelles.~\cite{Lammerding_2011_Mechanics_of_nucleus}  
The mechanics of the nucleus is determined by the interplay between the two nuclear 
membranes, the embedded nuclear pore complexes, the nuclear lamina, the nuclear cytoskeleton (which includes actin
filaments and myosin motors) and the different chromatin domains.
The combined effect of these factors leads to an effective nuclear stiffness that can be up to 10-fold stiffer 
than the rest of the cell,~\cite{Caille_2002_Nucleus_stiffness} which together with its size already 
suggests its importance in whole-cell mechanics. A very recent computational study showed
that even a spatially varying nuclear stiffness can be described on the whole cell level by
one effective modulus for the nucleus.~\cite{wohlrab2024mechanical} 

During recent years, it has been shown in many experimental studies that the nucleus indeed
has very specific mechanical roles in animal  cells. In matrix-driven cell differentiation, the nuclear stiffness correlates with tissue and matrix compliance, leading to stiffer cell nuclei on stiffer substrates and pointing at its ability of perceiving mechanical cues and adapting to them.~\cite{Swift_2013_nuclear_laminA_tissue_stiffness_matrix_directed_differentiation}
Recently, it has been demonstrated that nuclear deformations instruct migratory behaviour of cells 
in confined spaces, indicating that the nucleus serves as a ruler and mechanosensor.~\cite{Lomakin_2020_nucleus_acts_as_ruler_tailoring_cell_response_to_spatial_constraints,Venturini_2020_nucleus_measures_shape_changes} Moreover, nuclear size and stiffness limit the minimal size of constrictions through which a migratory cell can squeeze through.~\cite{Thiam_2016_Perinuc_Arp2_3_actin_polym_enables_nucl_deformation_to_faciitates_cell_migration} 
In turn, it has been observed that nuclear softening during passage of narrow constrictions is often associated with nuclear envelope rupture and DNA damage, which in our context are not only failure processes, but also signaling events.~\cite{Thiam_2016_Perinuc_Arp2_3_actin_polym_enables_nucl_deformation_to_faciitates_cell_migration,Denais_2016_nuclear_envelope_rupture_and_repair_during_cancer_cell_migration,Raab_2016_ESCRT3_repairs__nucl_envelope_ruptures_during_cell_migration}
Stresses and strains on the nucleus can also lead to structural changes in chromatin packing and a subsequent softening of the nucleus.~\cite{Wickstroem_2020_cell_heterochromatin_nuclear_softening_protects_genome_against_stresses}
It also has been shown that metastatic cancer cells use the nucleus as a "battering ram" to invade soft tissue.~\cite{Kristal_Muscal_2013} In cell migration, the nucleus is positioned by the microtubule-organizing center either at the front or the back, depending also on the properties of the environment; when
positioned at the front, it can be used as a ram during cell migration.
Last but not least, it is known that forces originating from the interplay between cytoskeleton and the cellular surrounding can be directly transmitted to the nuclear envelope leading to nuclear deformations, triggering transcriptional activities and cellular reactions to these stimuli. This direct mechanotransduction pathway includes the LINC protein complexes establishing a direct physical connection between nucleus and cytoskeleton.~\cite{Crisp_2006_firts_mention_of_LINC, Lombardi_2011_first_demo_of_LINC_function}

Despite this growing body of evidence of its importance for cell mechanics and mechanotransduction, the nucleus
is often neglected when modelling whole-cell mechanics, 
often due to lack of an appropriate theoretical framework.
We here propose an extension of our previously developed elastic phase field approach for cell mechanics~\cite{Chojowsi_2020_EPJE}
that also includes the nucleus. 
In the spirit of multi-phase field approaches,~\cite{Nonomura_multi_PF_2012,LoeberSR_2015,Wenzel_2021_multi_PF_cell_migration} the nucleus is introduced as an additional field, as was done in previous phase field studies of cells,\cite{Camley_micropattern_nucleus,Moure_Gomez_PF_nucleus} but
this time, we associate to the nucleus elastic material characteristics and make them different from the ones
of the rest of the cell. This enables us to study the effect of the nucleus 
on the cell's mechanical behaviour in a variety of different and biologically highly relevant situations,
including various boundary conditions between an adherent cell and the substrate as well as
compression and micropipette suction experiments of spherical cells. 
We also show that our approach is sufficiently general to allow for 
the investigation of viscoelasticity and cortical tension, which paves
the way towards more detailed models of nuclear mechanics in the future.

This work is structured as follows. First, we present the modelling approach for an elastic cell with a nucleus in section~\ref{secPFM}. We then demonstrate its applicability for homogeneously and locally adhered cells in section~\ref{Adhcell_const},
already pointing out an important role of the nucleus. For the simple geometry 
of an isotropically contracting, homogeneously adhering, disk-like cell with a nucleus,
we can use analytical solutions to 
validate the numerical solution.
We then proceed with discussing numerical studies of more complex experimental setups, 
namely patterned adhesion and dynamic failure of an adhesion point including viscoelastic relaxation.
In section~\ref{confined} we finally turn to cells in confinement and discuss
as examples the compression of cells between two parallel plates as well as 
micropipette aspiration. In this section we also study viscoelastic effects
of Kelvin-Voigt type.
We conclude with a discussion and outlook on possible applications and further extensions of the proposed method.

\section{Elastic phase field model 
for a cell with nucleus}
\label{secPFM}

To explicitly account for the cell's nucleus
in a model of an elastic cell 
in both stationary and dynamic situations, 
we extend the previously introduced elastic phase field approach.~\cite{Chojowsi_2020_EPJE}
The phase field method,
originally developed in the context of solidification processes~\cite{karma}
is nowadays widely used, especially
in the communities of fracture mechanics~\cite{Igor_crack,KarmaFracture} 
and poly-crystalline structures~\cite{polycrystal}.
Due to its ease in describing deformable or moving boundary problems,
applications spread out to soft matter physics,
e.g.~vesicles in flow~\cite{biben} or growing
actin gels~\cite{JohnMisbah_nonlin_sym_break}.
In the context of cellular biophysics, 
it proved efficient to model single cell migration~\cite{Shao2010,Ziebert2012,LoeberSM,Ziebert2016,Moure_Gomez_2018}  
and cell collectives,~\cite{Nonomura_multi_PF_2012,LoeberSR_2015,
Najem2016,Wenzel_2021_multi_PF_cell_migration}
as well as more recently cell~\cite{Chojowsi_2020_EPJE}
and tissue~\cite{YeomansPRL} mechanics.
Phase field models for a cell 
containing an explicit 
nucleus have been already proposed. 
However, Ref.~\cite{Camley_micropattern_nucleus}
neglected mechanics by solely considering the dynamics of two internal chemicals, while Ref.~\cite{Moure_Gomez_PF_nucleus} assumed Stokesian hydrodynamics.
To our knowledge, no phase field model has been proposed yet that would account for 
elastic continuum mechanics and
allow to model several cellular compartments -- 
here the cytoplasm and the nucleus --
having different material properties.

The study of moving boundary problems 
is a computationally expensive task because
at each point in time the location of the boundary has to be determined anew in order to impose the respective boundary conditions. 
The phase field approach circumvents
this problem by introducing 
an evolution equation for 
an auxiliary order parameter field  $\phi(\vec{x},t)$ (the phase field) describing the object of interest. 
It differentiates between two
bulk "phases", in our context the inside
of the object ($\phi = 1$) 
and its outside ($\phi = 0$), 
defined by the minima of a double-well potential.
Interfaces between these phases are then
given by smooth $\tanh$-like transitions 
from one bulk value to the other.
The location of the interface can be identified 
with the location of the maximum of the phase field gradient $|\nabla\phi|$, or simpler,
with the position of 
the isosurface with $\phi=1/2$.
If the evolution equation for the phase field
is coupled adequately 
to the other model equations
that describe the physical quantities of 
interest in the two phases
(e.g.~deformation, flow or diffusing chemicals),
the domain deforms and/or moves in response
to the processes described by these model equations.   

\begin{figure}[t!]
    \centering
    \includegraphics[width=\linewidth]{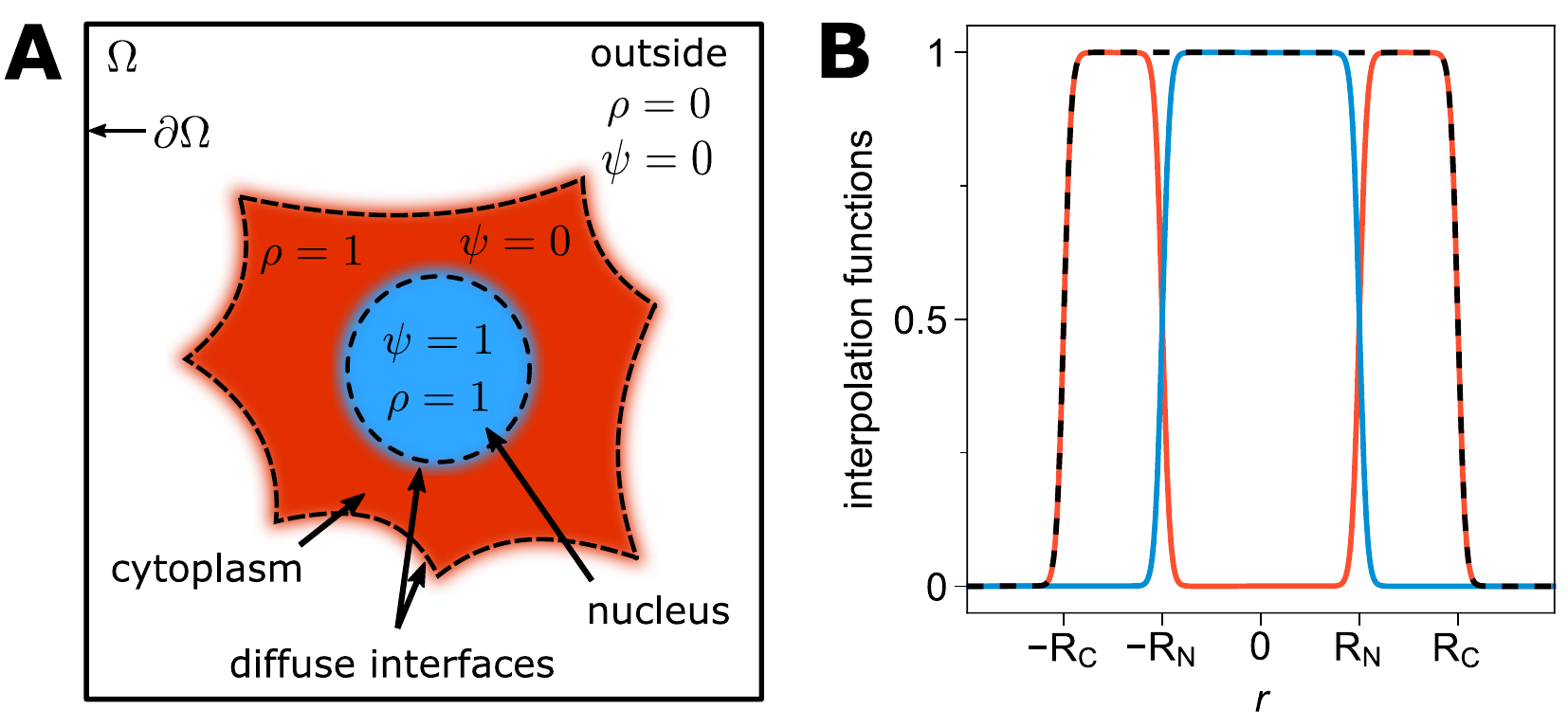}
    \caption{(A) Sketch of the two-phase field approach for modelling a cell with a nucleus. The computational domain $\Omega$ with boundary $\partial \Omega$ is divided into different compartments by use of the phase fields $\rho(\bm{x},t)$ and $\psi(\bm{x},t)$ for the whole cell and the nucleus, respectively. 
    The distinguished phases are the outside of the cell ($\rho=0, \psi=0$), the cytoplasm ($\rho=1, \psi=0$), and the nucleus, ($\rho=1, \psi=1$). (B) Radial cut showing the interpolation functions for a cell of diameter $2 R_C$ with a nucleus of diameter $2 R_N$. The cell ($h(\rho)$, black dashed line) is split into two compartments, the cytoplasm ($h(\rho)-h(\psi)$, red) and the nucleus ($h(\psi)$, blue).}
    \label{fig: figure 1}
\end{figure}

In our approach, the cell and its nucleus are represented by two phase fields, $\rho(\vec{x},t)$ and $\psi(\vec{x},t)$,
respectively, \textit{cf.}~Fig.~\ref{fig: figure 1}~A.  
Each field has its own evolution equation which follows an  overdamped relaxational dynamics
\begin{equation}\label{phase field eq. for rho}
    \partial_{t}\phi = D_{\phi} \Delta \phi - \partial_{\phi} g(\phi) + D_{\phi} \kappa_{\phi} |\nabla \phi| - \frac{1}{\xi} \left(\nabla \cdot \matr{\Sigma}+ \vec{F}_{tot} \right) \cdot \nabla \phi
\end{equation}
for $\phi\in\{\rho,\psi\}$, respectively. The first term penalizes the formation of interfaces whose width $\epsilon_\phi$ is set by the diffusion coefficient $D_\phi$ ($\epsilon_\phi \propto \sqrt{D_\phi}$). In general, the two interface widths could be chosen to be different. The second term is the derivative of a double-well potential of the form $g(\phi)=\phi^2(1-\phi)^2$.  Its minima are associated with the interior of the cell/the nucleus ($\phi=1$) and the space outside the cell/the nucleus ($\phi=0$), respectively, \textit{cf.}~again Fig.~\ref{fig: figure 1}~A. 
It should be noted that the exact form of the double-well potential is arbitrary; we opted for the 
simplest one. Inherent to the phase field approach is a wall energy (surface tension) that tends to pull together curved interfaces.~\cite{Allen_Cahn_1977,Folch99, Jamet_Misbah_2008_elim_surf_energy} Employing the third term in Eq.~(\ref{phase field eq. for rho}), proportional to the interface curvature $\kappa_\phi = -\nabla \cdot (\nabla \phi/|\nabla \phi|)$, allows to remedy this effect.~\cite{Folch99}  
Finally, we couple the phase field dynamics to the continuum mechanics via the last term in Eq.~\eqref{phase field eq. for rho}. 
It describes a movement of the phase field in case the mechanical force balance, 
$\nabla \cdot \matr{\Sigma}+ \vec{F}_{tot}=0$ with $\matr{\Sigma}$ being the stress tensor and $\vec{F}_{tot}$ all the forces acting on the domain, 
is not fulfilled.

The evolution of the displacement field $\vec{u}$ can be written, using the common assumption of overdamped dynamics for cells and tissues, as
\begin{equation}\label{elastodynamic eq.}
    \xi \partial_{t} \vec{u}=\nabla \cdot \matr{\Sigma} + 
    \vec{F}_{tot}.\
\end{equation}
Here $\xi$ sets the timescale of the relaxation into mechanical equilibrium,
given by the force balance. 
The total force,
$\vec{F}_{tot}=\vec{F}-\gamma(\vec{x})\left[1-h(\rho)\right]\vec{u}$,
contains all the applied forces 
$\vec{F}$ and a term that
suppresses artefacts in the displacement field that may arise in the outside phase due to reverting interface motions under force release. Eqs.~(\ref{phase field eq. for rho}), (\ref{elastodynamic eq.}) have been developed and verified in depth in Ref.~\cite{Chojowsi_2020_EPJE},
where more details can be found.

The stress tensor $\matr{\Sigma}$ has to be defined on the entire computational domain. 
In case of several compartments with different material properties, the phase field stress tensor has to interpolate the stress tensors $\matr{\sigma}$ (and lastly material parameters) of the individual considered phases, with  smooth transitions at the respective interfaces. For this purpose, we use weighting functions of the form $h(\phi)=\phi^2(3-2\phi)$ for the cell and the nucleus, respectively.~\cite{JohnMisbah_nonlin_sym_break,KassnerMisbah_stress_instab} The total phase field 
stress tensor $\matr{\Sigma}$ is then defined as 
\begin{equation}\label{phase field stress tensor}
    \matr{\Sigma}(\rho,\psi) = \left[h(\rho)-h(\psi)\right]\matr{\sigma^C} + h(\psi)\matr{\sigma^N}
\end{equation}
with $\matr{\sigma^{\bm{C}/\bm{N}}}$ being the stress tensors of the cytoplasmic ($\bm{C}$) (\textit{i.e.} the intracellular part without nucleus) and the nuclear compartment ($\bm{N})$.
The interpolation function for the cytoplasmic compartment is $h(\rho)-h(\psi)$(\textit{i.e.} cell, but not nucleus), \textit{cf.}~Fig.~\ref{fig: figure 1} B. 
As for the phase field potential, the form of the weighting functions is again not unique. They should, however, fulfil certain conditions, namely $h(1)=1$, $h(0)=0$ and $\partial_{\phi} h(1)=\partial_{\phi}h(0)=0$. 
Outside of the cell we assume the 
stress tensor to be zero for simplicity. 
Note that the cytoplasmic and the nuclear 
compartments are mechanically coupled (only) via the phase field stress tensor, Eq.~\eqref{phase field stress tensor}.

Finally, we have to specify the constitutive relation for the cytoplasm and the nucleus, respectively. We assume linear elasticity~\cite{landau_lifschitz} 
with the stress tensors defined as $\matr{\sigma}^\alpha=2\mu^\alpha\matr{\epsilon}+\lambda^\alpha tr(\matr{\epsilon})\matr{1}$, where $\alpha=\{C,N\}$ for cytoplasm (C) and nucleus (N). Here, $\mu^\alpha$ and $\lambda^\alpha$ are the Lam\'{e} coefficients of each compartment. The strain tensor $\matr{\epsilon}$ is defined in index notation as $\epsilon_{ij}=(1/2)(\partial u_i/\partial x_j + \partial u_j/\partial x_i)$ and $\matr{1}$ is the identity matrix. 

In three-dimensions, 
the Lam\'{e} coefficients are given by $\lambda_{3D}=\nu E /[(1+\nu)(1-2\nu)]$ and $\mu_{3D}=E/[2(1+\nu)]$ with Young's modulus $E$ and Poisson's ratio $\nu$.
Depending on the geometry of the considered problem, different two-dimensional approximations can be used:
Strongly spread cells, having a 
height $d$ (assumed to be along the $z$-axis) considerably smaller than the lateral extensions, can be approximated as thin elastic sheets in plane stress formulation. 
In this case the stress components 
$\sigma_{zz}=\sigma_{xz}=\sigma_{yz}=0$ vanish and the problem becomes effectively 
two-dimensional with thickness-averaged $\lambda_{2D}=\nu E d/(1-\nu^2)$ and $\mu_{2D}=Ed/[2(1+\nu)]$.~\cite{Howell} 
For a cell having the shape of a 
long cylinder (again in $z$-direction), 
the plane strain formulation can be applied, where $\epsilon_{zz}=\epsilon_{xz}=\epsilon_{yz}=0$.~\cite{Howell} Here, the Lam\'{e} coefficients are identical to the three-dimensional ones. We will specifically mention the used approximation for each experiment discussed in the following.

\section{Modeling spread cells}
\label{Adhcell_const}

We now demonstrate the applicability of the proposed method by investigating a cell 
of height $d$ spread onto a compliant substrate 
in 2D plane stress formulation. 
This situation is biologically highly important,
since cells are able to sense the mechanical properties of their environment via internal force generation and transmission of these forces to the outside.~\cite{schwarz_soft_2007}
The received information can then be used by the cell to adapt
its mechanical properties and morphology, 
and possibly even to induce division, differentiation or motility (processes which are beyond the scope of this work).
To model a spread cell, 
we have to include active cell contractility 
and cell-substrate adhesion as central features into the proposed method.

\begin{figure*}[t!]
    \centering
    \includegraphics[width=\linewidth]{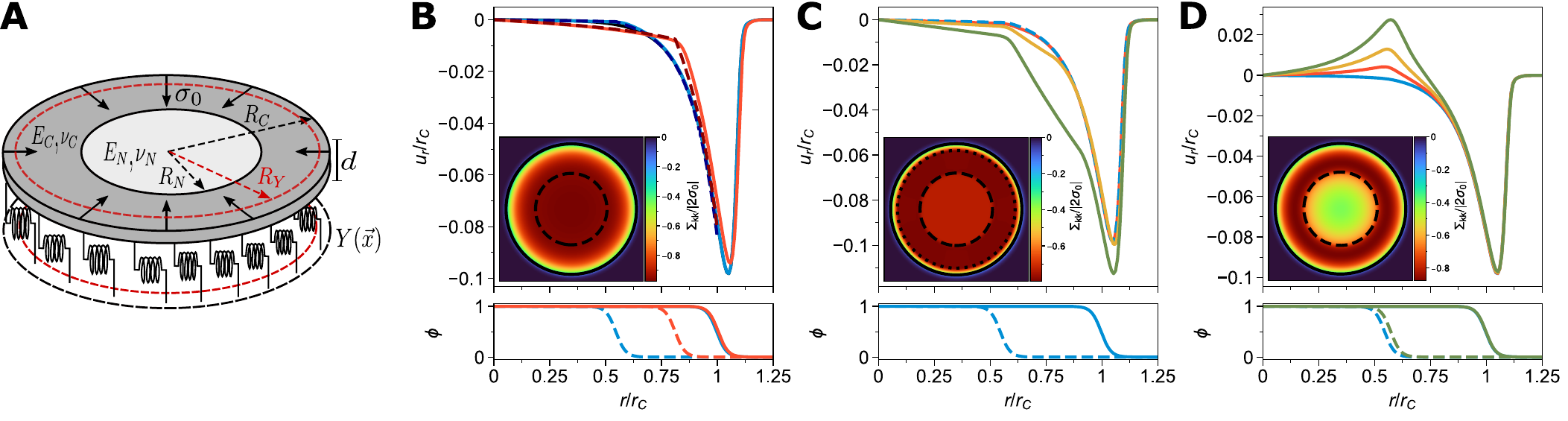}
    \caption{
    (A) 
    Sketch of the model for a cell (thickness $d$, radius $R_C$) with a concentric nucleus (radius $R_N$). The cell is contracting isotropically with active stress $\sigma_0$ while being adhered to a substrate via a spring stiffness density $Y(\vec{x})$ on a ring 
    $R_Y \leq r \leq R_C$. 
    The cytoplasm ($E_C$, $\nu_C$) and the nucleus ($E_N$, $\nu_N$) can have different material properties.
    (B) The homogeneous adhesion case
    with $R_Y=0$. Shown in the upper panel are the
    phase field (solid) and analytical (dashed)
    solutions for the radial displacement field $u_r$, normalized by the cell radius in mechanical equilibrium, $r_C$. Shown are the cases: $E_N/E_C=10$, $R_N/R_C=0.5$ (blue); $E_N/E_C=10$,   $R_N/R_C=0.75$ (red); and the phase field solution for $E_N/E_C=1$ and $R_N/R_C=0.5$ (black, mostly covered by the blue curve). 
    The inset shows the trace of the stress tensor, normalized by the active stress $\sigma_0$, for $E_N/E_C=10$, $R_N/R_C=0.5$; the contour lines correspond to $\rho=0.5$ (cell, solid) and $\psi=0.5$ (nucleus, dashed), The lower panel shows the radial profile of $\rho$ (cell, solid) and $\psi$ (nucleus, dashed) in mechanical equilibrium with colors corresponding to the upper panel. 
    (C) Adhesion on an outer ring only. In the upper panel phase field solutions for $u_r/r_C$ are shown for
    $E_N/E_C=10$, $R_N/R_C=0.5$
    and varying
    $R_Y/R_C=0.5,0.6,0.7.0.8$ (blue to green). 
    The inset shows the normalized stress for the case $R_Y/R_C=0.8$ (dotted line marks inner ring boundary) and the lower panel the phase field profiles for the case $R_Y/R_C=0.5$ (blue).  
    (D) Until now, the case $f=0$ was considered (\textit{cf.}~Eq.~(\ref{fdef})) \textit{i.e.}~the whole cell was contracting.
    When gradually restricting contraction to the cytoplasm
    by varying $f=0,0.2,0.5,1$ 
    (blue to green), cf.~the discussion 
    in the main text,
    extensile forces are exerted
    on the nucleus of radius $R_N/R_C=0.5$. The lower panel shows the phase field profiles for
    $f=0$ (blue) and $f=1$ (green).
    The inset shows the normalized stress for the case 
    $E_N/E_C=10$ and $f=1$, 
    for better comparison with the insets of B and C. 
    All simulations were performed on $N=512\times512$ grid points on a domain of $50\,\mu{\rm m} \times 50\,\mu{\rm m}$. 
    If not specified above, the other mechanical parameters are 
    $R_C=12.5\,\mu{\rm m}$, $d=1\,\mu{\rm m}$, $E_C=\sigma_0=1\,{\rm kPa}$, $\nu_C=\nu_N=0.5$ and $Y_0=0.8\,{\rm nN}/\mu{\rm m}^{3}$. Further parameters are as in table~\ref{tbl:parameters}.
    }
    \label{fig:figure2}
\end{figure*}

Active stresses $\matr{\Sigma_{act}}$ can be straightforwardly introduced into the phase field stress tensor, Eq.~\eqref{phase field stress tensor}, as an additive contribution.
In principle, the active stress can be time- 
and space-dependent. 
Contractile stresses within a cell arise due to the activity of myosin II motor proteins, which slide cytoskeletal actin filaments relatively to each other.~\cite{Howard_2001} 
While some part of the contracting cytoskeleton spans over the nucleus, 
other parts can also bind directly to 
it via LINC complexes,  exerting 
contractile stress on the nuclear boundary.~\cite{Crisp_2006_firts_mention_of_LINC, Lombardi_2011_first_demo_of_LINC_function} 
Using the common approximation of an isotropic contractile stress $\matr{\sigma}_{act}=\sigma_0 d \matr{1}$, with $\sigma_0>0$ and 
$\matr{1}$ the identity matrix, we write the active stress tensor as
\begin{equation}
\label{fdef}
\matr{\Sigma_{act}}=[h(\rho)-fh(\psi)]\sigma_0 d \matr{1}.    
\end{equation}
The function in the bracket indicates 
in which cell compartment the 
active contractile stress is acting.
We consider two extreme cases which represent processes in the third dimension that
here are integrated out. 
For $f=0$, the whole cell is 
under contractile stress, 
including the nucleus.
Hence only the forces at the cell 
boundary are unbalanced and 
effectively contract the cell.
Such a situation can result from the
presence of a strong perinuclear actin cap.~\cite{Wirtz_actincap} 
For $f=1$, only the cytoplasm 
is contracting.
Then the nucleus 
effectively feels an extensile force.
This situation should result if 
the complete contractile apparatus is
localized in the cytoplasm and does not
bridge over the nucleus. By varying
$f$ from $0$ to $1$, we can tune the
strength of this effect. 
In the following, we consider 
$\matr{\sigma}_{act}$ to be 
time-independent and homogeneous in 
the respective cell compartments
and investigate only 
steady state situations.

The second feature needed to model spread cells
is cell-substrate adhesion, anchoring 
the cell and allowing for force transmission from the cytoskeleton to the substrate via integrin-mediated adhesion sites. 
A simple approximation for a fully elastic substrate
is an elastic foundation,
where adhesion sites are modeled as a spring stiffness density $Y(\vec{x})$ resisting cell deformations.~\cite{Edwards_2006,EdwardsSchwarz,banerjee_contractile_2012}
The associated restoring force entering the elastic Eq.~\eqref{elastodynamic eq.} is then given by
\begin{equation}\label{adhesion force}
    \vec{F}(\vec{x})= - Y(\vec{x})h(\rho)\vec{u} 
\end{equation}
where $h(\rho)$ indicates that adhesion sites can only form underneath the cell.
In principle, $Y(\vec{x})$ could be made time-dependent as well, allowing to model dynamics of bond formation.

\subsection{Adhering cell with radial symmetry}

We first study  a circular cell 
which is spread and actively contracting on an elastic foundation, 
as shown in Fig.~\ref{fig:figure2} A.
This geometry was originally used to explain the experimentally observed concentration of traction forces at the cell periphery from a mechanical perspective
and is analytically solvable for homogeneously adhered cells.~\cite{EdwardsSchwarz,banerjee_contractile_2012,Schwarz_Safran_2013}  
Recently, an analytical solution for the case where adhesion is restricted 
to a ring 
at the cell's periphery has been
also given.~\cite{Solowiej-Wedderburn_Dunlop_2022_cell_adhesion_pattern}
To benchmark our numerical framework, we generalized the homogeneous adhesion model
by additionally considering 
a disk-like nucleus in the cell's center; the analytical solution 
is given in Appendix~B. 
To specify the different possible geometries, we introduce
the cell's radius $R_C$, the nucleus' radius $R_N$ and the radius of the 
adhesive ring $R_Y$, meaning that the cell adheres for $R_Y\leq r \leq R_C$.

We begin with the simplest case of a cell fully and homogeneously adhered to the substrate, \textit{ i.e.}~$Y(\vec{x})=Y_0$
and $R_Y \to 0$, and 
assume the contracting cytoskeleton spans over the nucleus,  $f=0$. Fig.~\ref{fig:figure2} B upper panel shows the phase field (solid curves) 
and analytical solutions (dashed) 
for the radial displacement field $u_r$ for different nuclear stiffnesses and radii. 
Both are in very good agreement, confirming our approach. 
Deviations result from the diffuse description of the nucleus-cytoplasm boundary in the phase field framework
and can be reduced by decreasing its interface width. 
The kink at the nucleus-cytoplasm interface, occurring in both the analytical and numerical solution, 
is due to the different rigidities of the two considered cell compartments. 
Consistent with previous results, the highest deformations are at the cell periphery.~\cite{EdwardsSchwarz}
This is associated with high traction stresses
at the periphery and lowered total internal stresses, as visualized
in the inset of Fig.~\ref{fig:figure2} B
by plotting the trace of the stress tensor, 
normalized by the active stress level $\sigma_0$.

How important is the nucleus for the mechanics?
For a nucleus of half the cell's radius, $R_N/R_C=0.5$, the nucleus stiffness $E_N$ has only a negligible effect on the cell's deformation. 
Increasing the nuclear radius,
a realistically stiff ($E_N/E_C=10$) nucleus 
(red curves) leads to considerably different  slopes in the displacement field. 
However, the overall position of the cell periphery remains approximately the same, \textit{cf.}~the solid curves in the lower panel of Fig.~\ref{fig:figure2} B, displaying the radial phase field profiles.

It is important to note that the displacement field in the nucleus always remains small.
This demonstrates that strong cell adhesion protects
the nucleus against large deformations and stresses. 
The determining factors are the distance between the nucleus and the cell boundary, $R_C-R_N$, 
and the characteristic distance over which 
stress can propagate through the cytoplasm, 
which for an adhering cell is given by the localization length $l_C=\sqrt{E_Cd/Y(1-\nu_C^2)}$.~\cite{EdwardsSchwarz}
Peripheral cell adhesion is sufficient for protecting the nucleus, corresponding to 
the experimental observation of strong adhesions
forming mostly at the cell periphery, 
while the basal side under the nucleus is mostly adhesion free.~\cite{zamir2001molecular} 

This shielding can have a major impact on the nuclear mechanosensing ability of stimuli originating at the cell edge.
In a second study we therefore restrict the adhesion to a ring at the cell periphery of inner radius $R_Y$, to see whether the nuclear deformation increases, indicating a higher perception of mechanical stimuli. 
In most cell types, the nucleus occupies 
not more than a third of the cellular volume. Therefore, we fix the nucleus radius to $R_N/R_C=0.5$, 
for which we found above that the nuclear stiffness has only a negligible effect on cell mechanics, and the stiffness to $E_N/E_C=10$.
We then examine the radial deformation upon varying the inner radius $R_Y$ of the adhesion ring as shown in the upper panel of Fig.~\ref{fig:figure2} C. 
Note that the deformation field is linear in the non-adhered cell parts, \textit{i.e.}~both in the nucleus and the inner part of the cytoplasm. 
As visible from the displacement field, 
a larger $R_Y$, and therefore a decreased 
adhesion area, increases the deformation the nucleus experiences. This demonstrates 
that adhesion restricted to the cell periphery leads to an increased stress propagation to the nucleus, as also visible in the inset of Fig.~\ref{fig:figure2} C. Nevertheless large deformations are prevented 
as shown by the only slightly increased deformation peak compared to the fully adhered case in Fig.~\ref{fig:figure2} B upper panel. 
This agrees with recent experiments on optogenetic
activation of whole cells that showed that disc and ring
geometries give little differences in regard to
whole-cell contractility.~\cite{andersen_cell_2023}

Lastly, we study the situation of a fully adhered disk, but now with a varying parameter $f$ as described above. 
Increasing $f$ from 0 to 1,
leads to an extensile stress 
on the nuclear boundary.
Fig.~\ref{fig:figure2} D upper panel demonstrates the radial displacement field for $R_N/R_C=0.5$ and $E_N/E_C=2$ for different 
$f=0,0.2,0.5,1$, 
the peaks close to the nucleus-cytoplasm interface clearly showing a radial stretching of the nucleus,
which is also visible in the lower panel of Fig.~\ref{fig:figure2} D showing the phase field profiles for the cases $f=0$ (blue) and $f=1$ (green). Similar observations can be made for other nuclear rigidities. For increasing 
parameter $f$, the nucleus experiences higher extensile stresses, also visualized in the inset of Fig.~\ref{fig:figure2}, in contrast to the previous discussed cases.

In summary, the above results verify our elastic phase field approach
and indicate that the transmission of mechanical cues to the nucleus strongly depends 
on the actual adhesion geometry and the 
force transmission from the cytoskeleton to the nucleus.

\begin{figure}[t!]
    \centering
    \includegraphics[width=\linewidth]{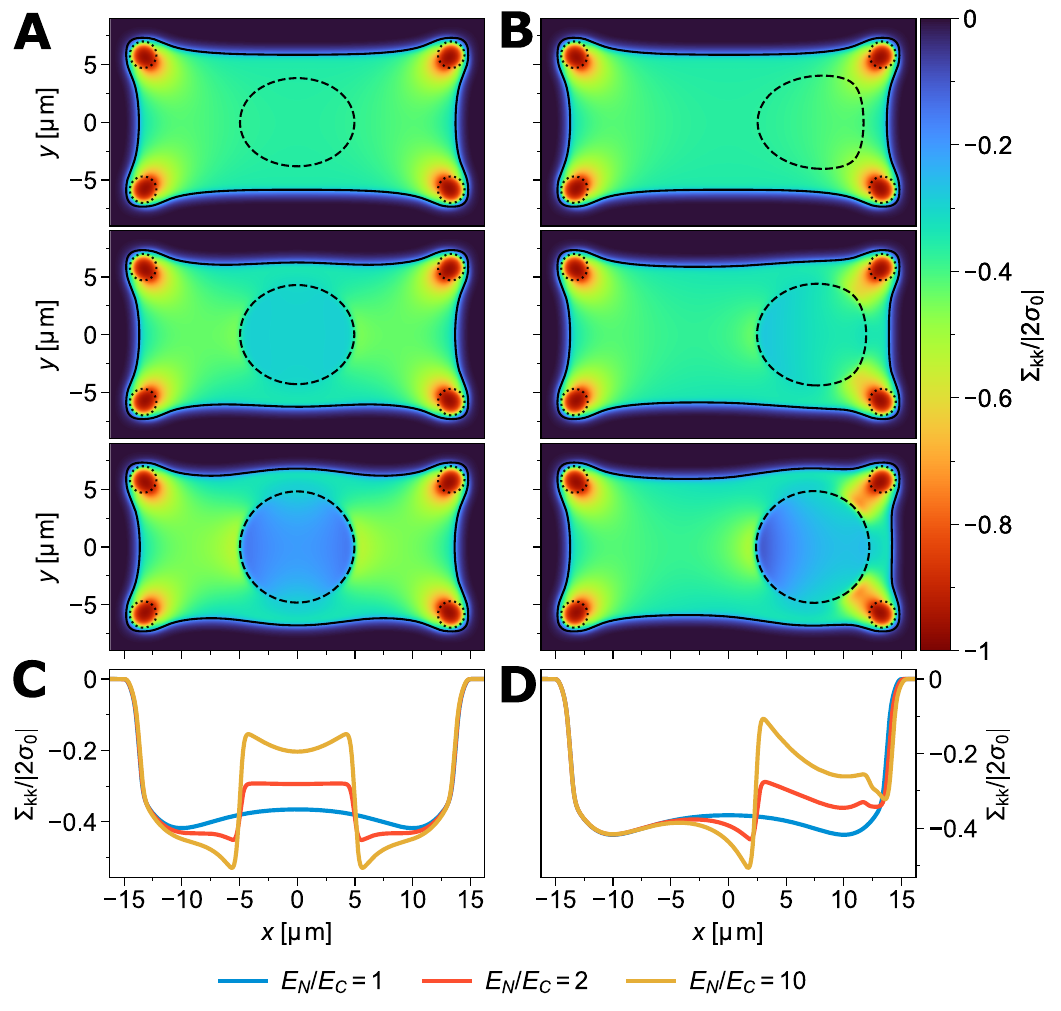}
    \caption{A cell adhering in a rectangular shape due to adhesive spots in the corners. Adhesive spots (dotted) have a radius $r_{adh}=1.15\,{\rm \mu m}$ with high $Y_0=16\,{\rm nN/\mu m^3}$, to prevent slipping from the adhesion sites.
    The cell contracts under an isotropic contractile stress $\sigma_0/E_C=0.4$. 
    Shown is the normalized trace of the stress tensor for the cases $E_N/E_C=1,2,10$ (top to bottom) with an initially circular nucleus (A) centered in the cell and (B) shifted in $x$-direction by $1.5 R_N$. The contour lines correspond to $\rho=0.5$ (solid, cell) and $\psi=0.5$ (dashed, nucleus). 
    (C) and (D) show the trace of the stress tensor along the symmetry line $y=0$ for the corresponding simulations shown in (A) and (D). 
    All simulations were performed on $N=1024 \times 512$ grid points on a domain of $50\,\mu {\rm m}\times 25\,\mu {\rm m}$. Initial cell dimensions are $30\,\mu {\rm m}\times15\,\mu {\rm m}$ with $R_N= 5\mu m$, $d=1\,\mu{\rm m}$, $E_C=1\,{\rm kPa}$ and $\nu_C=\nu_N=0.5$. Rest as in table~\ref{tbl:parameters}.}
    \label{fig:figure3}
\end{figure}

\subsection{Contractile cells on adhesion patterns}

Micro-patterned adhesive substrates are a standard setup for studying cellular behaviour in structured environments.~\cite{Chen_1997,Thery_2006,Lehnert_2004,link2023cell}
Adherent cells are always under contraction,
as nicely demonstrated by the ubiquitous invaginated
arcs that form when cells adhere with point-like adhesions.~\cite{uss:bisc08a,Schwarz_Safran_2013}
Here, we investigate the impact of the nucleus on the overall cell morphology in such geometries.
As a first example, we study a rectangular pattern with
four circular adhesive patches
of radius $r_{adh}$ located at its corners. We start with a 2D rectangular cell,
described in plane stress, 
and allow it to form focal adhesions at the corners and contracting isotropically under a  contractile stress $\sigma_0$. The nucleus initially has a circular shape of radius $R_N$ with physiological nucleus-to-cell volume ratio $V_N/V_C\approx0.17$. 
We consider the case 
that the cytoskeleton contracts 
the whole cell, \textit{i.e.}~$f=0$.
For the adhesion strength $Y(\vec{x})$ we use a smoothly varying field, transitioning in a tanh-like manner from the maximal value $Y_0$ in the focal adhesion towards zero outside of it.
Primarily, this ensures numerical stability compared to pinning the cell completely to the focal adhesion (via the boundary condition $\vec{u}=0$, \textit{cf.}~also Ref.~\cite{Chojowsi_2020_EPJE}). 
It also would allow to study different adhesive strengths in different focal adhesions.

Representative results are shown in
Fig.~\ref{fig:figure3}.
The panels of Fig.~\ref{fig:figure3} A
study a centered nucleus and 
demonstrate the effect of an increased nuclear-cytoplasmic stiffness ratio $E_N/E_C$. Clearly, the nucleus is deformed by the 
invaginated arcs for low nuclear stiffness. A higher nuclear stiffness rather changes the shape of the cell,
demonstrating again
that localized adhesion and an increased nuclear stiffness protect the nucleus against large deformations/stresses.
Yet one also sees how stress bridges start to emerge between nucleus and adhesions, which
look like precursors of stress fibers. 
Similar perturbations are observable for example for cells spreading on nanonets.~\cite{Jana_2022_Sculpting_rupture_free_nuclear_shapes_in_fibrous_envirnoment}  

Similar to the previous study in circular geometry, 
the distance between the nucleus and the cell edge is a determining factor for the magnitude of the morphology perturbation. 
If the nucleus position is shifted away from the cell's center, \textit{cf.}~Fig.~\ref{fig:figure3}~B, 
a stress accumulation at the cytoplasm-nucleus boundary can be observed, while the stress is lowered
on the opposite side of the nucleus.
In Fig.~\ref{fig:figure3} C and D the trace of the stress along the symmetry line $y=0$ is depicted, clearly showing the stress decrease for higher $E_N/E_C$ and its asymmetry when shifting the nucleus.
Interestingly, as visible in Fig.~\ref{fig:figure3} B,
the stress "builds a bridge" 
between the closeby focal adhesions and the nucleus,~\cite{Ronan_2014} quite possibly impacting the mechanosensing of the nucleus.
Furthermore, one can hypothesize 
that the asymmetric stress distribution for shifted nuclei allows the cell to differentiate between left and right, which may be important
to polarize for cell migration.

\subsection{Failure of a focal adhesion}
\label{failadh}

\begin{figure}[t!]
    \centering
    \includegraphics[width=\linewidth]{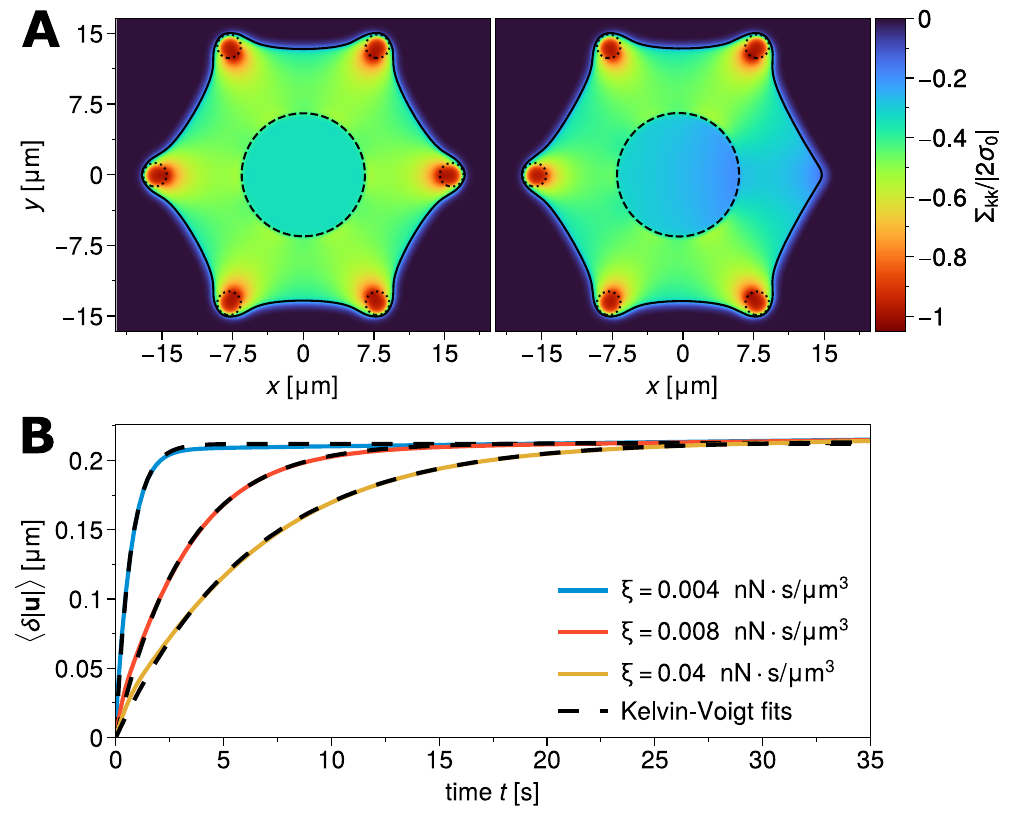}
    \caption{(A) A cell with nucleus 
    was allowed to spread in a hexagonal adhesion pattern and to contract isotropically with $\sigma_0/E_C=0.4$ until it reached mechanical equilibrium (left panel). Subsequently the most right adhesion spot was removed and the cell evolved towards a new mechanical equilibrium (right panel). The colormap shows the normalized trace of the stress tensor.  The cell shape (isocline $\rho=0.5$, solid black) and nucleus shape ($\psi=0.5$, dashed black) are also shown. 
    (B) Shown is the average displacement
    $\langle \delta|\vec{u}|\rangle$,
    with respect to the initial reference displacement in (A),
    as a function of time $t$ for different friction coefficients $\xi$. 
    For all tested $\xi$, the behavior is the one of a  
    Kelvin-Voigt model. 
    The simulations were performed on $N=512\times512$ grid points on a domain of $50\,\mu {\rm m}\times 50\,\mu {\rm m}$. Initial cell edge length is $17.5\, \mu{\rm m}$ and $R_N=6.65\,\mu{\rm m}$ with cell height $d=1\,\mu{\rm m}$ resulting in $V_N/V_C\approx0.17$. Further, 
    $E_N/E_C=10$ with $E_C=1\,{\rm kPa}$, $\nu_C=\nu_N=0.5$, $r_{adh}=1.25\, \mu{\rm m}$ and $Y_0=16\, {\rm nN/\mu m^3}$. 
    Rest as in table~\ref{tbl:parameters}.
    }
    \label{fig:figure4}
\end{figure}

Having demonstrated that the 
proposed modeling framework 
is able to describe static 
spread elastic cells with nucleus in 
complex geometries, 
we now give an example of a simple 
dynamic response. 
Similar to the last example, we 
consider a cell on a micro-patterned
adhesive environment favoring a 
hexagonal cell shape. 
The cell first contracts isotropically under a stress $\sigma_0$ until it reaches mechanical equilibrium.
The resulting shape, including stress focusing at the adhesion spots and invaginated arcs in between,
is shown in the left panel of Fig.~\ref{fig:figure4} A.
Afterwards, one of the adhesion spots 
(here, the most right one) is suddenly removed, mimicking the rupture/dissolution of a focal adhesion, and the cell deforms into a new mechanical equilibrium
given by this geometry, 
see the right panel of 
Fig.~\ref{fig:figure4} A. 
One can clearly see that the cell relaxes 
an substantial amount of stress in the area
of the missing adhesion point. 
The stress inside the nucleus is also reduced, in the shown example 
by $14.5$ \%, 
and again shows an asymmetry.
Note that the cell does not fully round up in the region close to the 
detached adhesion point, which is a consequence of the
reference state of the elastic model.

To quantify the dynamics of this relaxation,
we investigated 
the cell-averaged displacement 
$ \langle|\vec{u}|\rangle = (1/V_{cell}) \int \rho |\vec{u}| d\Omega$, 
where the cell's volume is given by $V_{cell}=\int \rho d\Omega$.
Fig.~\ref{fig:figure4} B shows
$\langle \delta|\vec{u}|\rangle=\langle|\vec{u}|\rangle-\langle|\vec{u_{ref}}|\rangle$, \textit{i.e.}~the deviation from the reference displacement
at the time point of the removal of the focal adhesion, as a function of time and
for different friction coefficients $\xi$ (\textit{cf.}~Eq.~\eqref{elastodynamic eq.}) 
and $E_N/E_C=10$. 
As can be seen, the displacement 
$\langle \delta|\vec{u}|\rangle$
always levels at the same plateau value, reflecting that mechanical equilibrium is reached, 
with $\xi$ determining the relaxation time.

It should be noted that in Ref.~\cite{Chojowsi_2020_EPJE}
the elasto-dynamic formulation 
of Eq.~\eqref{elastodynamic eq.} 
was introduced  out of necessity 
to couple the phase field dynamics with  elasticity in a reversible fashion. 
Hence, if one wants to describe a system with "pure" elastic behavior, one should not probe the system on time scales $\tau$ faster then the one set by $\xi$.
On the flip side, if one does so,
the average displacement follows the relaxation behaviour of a viscoelastic material with long-term elastic behavior.
This is reflected by the dashed curves in 
Fig.~\ref{fig:figure4} B
where we applied a Kelvin-Voigt model,
predicting
$\langle\delta|\vec{u}(t)|\rangle= 
u_{max}[1-\exp(-t/\tau_{R})]$,
to interpret the data, which fits perfectly.
Here $u_{max}$ is the maximum average displacement and $\tau_{R}$ the characteristic relaxation timescale. 
The Kelvin-Voigt model is a widely used and experimentally validated model for cellular mechanics, describing that mechanical relaxation does not occur instantaneously (as in linear elasticity), but is retarded by
internal friction, stemming from viscous flow and cytoskeletal reorganization.  
For a Kelvin-Voigt material the relaxation timescale is given by $\tau_{R}=\eta/E$,
where $E$ is the Young's modulus and $\eta$ the material's viscosity.
We verified that the correspondence to a Kelvin-Voigt model holds for all tested nucleus stiffnesses and the above comparison hence allows to associate $\xi$ with an effective viscosity $\eta$. 
Note that, however, since the cell
is a composite material of cytoplasm and nucleus, both $E$ and $\eta$ entering $\tau_R$ are cell-averaged quantities.
We will revisit viscoelastic effects
in section \ref{comp_VE}.

\section{Cells in confinement and modulus measurements}
\label{confined}
We now turn to the problem of cells in confinement, again focusing on the effects of the nucleus. 
On the one hand, in their physiological environment, cells are often subject to (dynamic) straining induced by their surrounding. 
Examples include cyclic stretching in lung and vascular tissue 
or the migration of immune cells and metastatic cancer cells through narrow openings in tissues or fibrous networks.
On the other hand, several experimental methods have been developed to probe cellular mechanical responses, including
compression of cells between 
two plates\cite{Yoneda_1964_compression,Thoumine_1997_compression,Ofek_2009_compression,Mitrossilis_2009_compression,Fischer-Friedrich_2016_rheology_of_active_cell_cortex_in_mitosis,Lomakin_2020_nucleus_acts_as_ruler_tailoring_cell_response_to_spatial_constraints,Venturini_2020_nucleus_measures_shape_changes} 
and cell aspiration by micropipettes.~\cite{Evans_Yeung_1989,Chesla_1998_micropipette,Evans_2004_micropipette,Herant_Dembo_2005_micropipette,Tinevez_Paluch_2009_micropipette,Robinson_2013_micropipette}
Here, we show how the latter 
two can be modeled using our framework to extract effective
elastic moduli.

\begin{figure*}[t!]
    \centering
    \includegraphics[width=\linewidth]{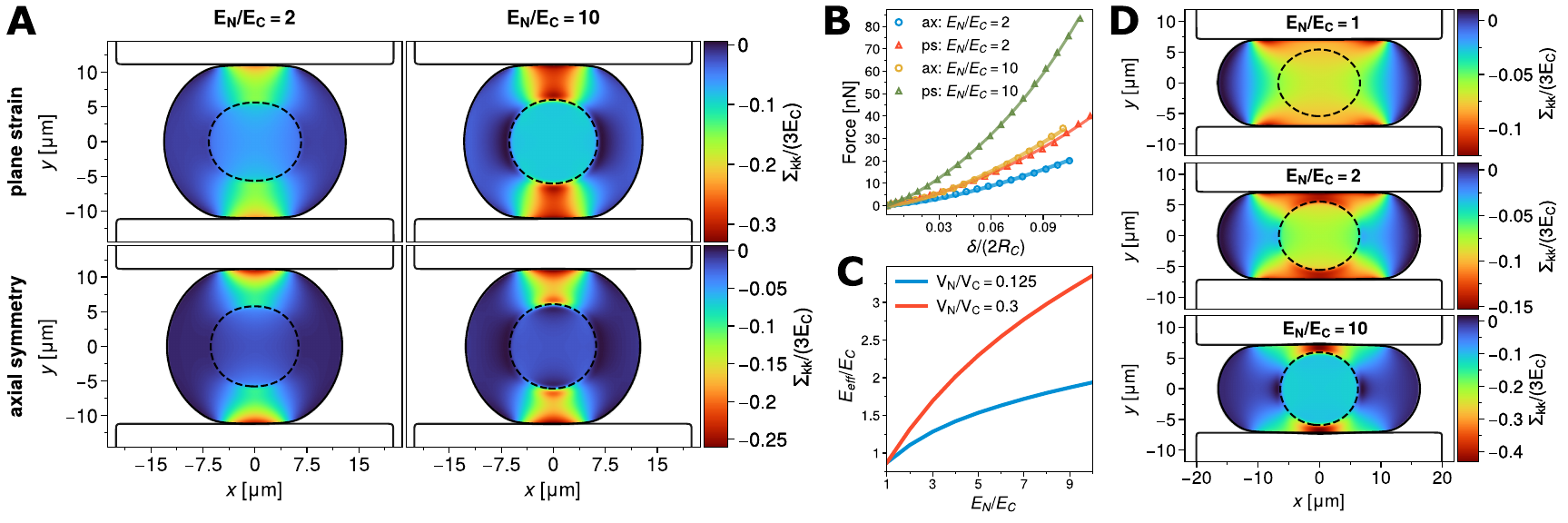}
    \caption{
    (A) Compression of a cell in the plane strain geometry (long cylinder, top) compared to a spherical cell (axial symmetry, bottom). Shown are the cases
    $E_N/E_C=2$ and $E_N/E_C=10$.
    (B) Numerically obtained force-compression curves. The symbols are numerical solutions with plane strain (triangles) and axial symmetry (circles), respectively, \textit{cf.}~panel A. 
    The solid curves are fits to the respective analytical solutions (available in the absence of the nucleus).
    (C) For the case of axial symmetry, we extracted an effective elastic modulus from fits as shown in panel B.
    In the physiological range of nucleus sizes and stiffness,
    the effective modulus measured in compression is up to three times larger than the one of the pure cytoplasmic stiffness.
    Colors in B: 
    $E_N/E_C=2$ in axial symmetry (blue), $E_N/E_C=10$ in axial symmetry (yellow); 
    $E_N/E_C=2$ in plane strain (red), $E_N/E_C=10$ in plane strain (green).
    Colors in C: nucleus size of $V_N/V_C=0.125$ (blue);
    nucleus size $V_N/V_C=0.3$ (red).
    (D) Compression experiment similar to (A) but for an axially symmetric, pancake-shaped cell.
    Simulations for (A)-(C) were performed on $N=512\times512$ grid points and for (D) on $N=512\times256$ grid points.
    Mechanically relevant parameters for all shown simulations (if not mentioned otherwise) are  $R_N=6.25\, \mu{\rm m}$ in (A) and (D),
    $E_C=1\, {\rm kPa}$, $\nu_C=\nu_N=0.48$, and $\alpha=6\, {\rm kPa}$. Rest as in table~\ref{tbl:parameters}.
    }
    \label{fig:figure5}
\end{figure*}

In both experimental setups, the interaction of the examined cell with the confining obstacles -- the plates of the compression apparatus or the tube walls of the micropipette -- is crucial. 
In the phase field method, such "obstacles" can be described by implementing another, static
phase field $\varphi(\vec{x})$,
also having tanh-like transitions from $\varphi=1$ within the obstacle to $\varphi=0$ outside, and
which is assumed here to be perfectly rigid. 
The local presence of the obstacle 
is then manifesting itself by interactions
of the cell's phase field with $\varphi$. 
Motivated by a phenomenological excluded volume potential of the form $\mathcal{F}=\frac{\alpha}{2}\rho^2 \varphi^2$ presented earlier,~\cite{Nonomura_multi_PF_2012,LoeberSR_2015,Ziebert2016} we add the following excluded volume force 
to the force $\mathbf{F}_{tot}$
entering Eqs.~(\ref{phase field eq. for rho}), (\ref{elastodynamic eq.}):
\begin{equation}\label{volume exclusion force}
    \vec{F}_{excl} = \alpha\rho \varphi^2 
    \frac{\nabla h(\rho)}{f(h(\rho))} \,. 
\end{equation}
Here the first term, including the interaction strength $\alpha$,
is the derivative of the excluded volume energy. $\nabla h(\rho)$ indicates
that the volume exclusion force acts orthogonal to the $\rho$-interface and is restricted to the interface region.
Finally, $f(x)=\sqrt{1+\epsilon(\nabla x)^2}$ with a small $\epsilon \ll 1$ implements saturation of the force
in case the phase field gradient 
becomes too steep.~\cite{LoeberSR_2015}

\subsection{Compressing cells between two parallel plates}
Compressing cells between two parallel plates is nowadays a standard experimental technique
to mechanically probe global cell mechanics.~\cite{Thoumine_1997_compression,Caille_2002_Nucleus_stiffness,Desprat_2006_compression,Ofek_2009_compression,Mitrossilis_2009_compression,Fischer-Friedrich_2016_rheology_of_active_cell_cortex_in_mitosis,Mokbel_Fischer_Friedrich_2020_compression_Poisson}
For instance, in combination with computational predictions, it has been demonstrated that for mitotic cells the cell cortex dominates cell mechanics.~\cite{Fischer-Friedrich_2016_rheology_of_active_cell_cortex_in_mitosis}
Beyond that, also cellular responses to increased confinement have been addressed, evidencing that it can induce the mesenchymal-amoeboid transition~\cite{Liu_Piel_Cell_2015}
and trigger cell migration.~\cite{Lomakin_2020_nucleus_acts_as_ruler_tailoring_cell_response_to_spatial_constraints, Venturini_2020_nucleus_measures_shape_changes}
In the latter studies it was suggested that the extent of nuclear compression
determines the onset of this response. 
Also a recent computational study
has shown that a stiff nucleus increases 
the effective stiffness of cells 
as probed in such experiments.~\cite{wohlrab2024mechanical}

We model compression experiments by implementing the upper and lower plates via the field $\varphi(\vec{x})$.
Both plates are initially not in contact with the cell, such that $\vec{F}_{excl}=0$. 
They are moved towards each other 
successively by the grid spacing $\Delta x$ 
each time the cell has relaxed into mechanical equilibrium. 
Having reached the desired compression level/plate distance, this procedure can be reversed to release the cell from the confinement. 
Note that we study the 
quasi-static, purely elastic  process 
first, to be able to compare 
with analytical solutions. 
Cell compression that is continuous 
in time,
where the response will then be of 
Kelvin-Voigt-type, \textit{cf.}~section~\ref{failadh}, will be investigated
in section \ref{comp_VE}.

So far, in section~\ref{Adhcell_const} 
we used an effectively 2D 
plane stress approach, which was justified for a thin, spread cell. 
In the compression experiment, 
the simplest effective 2D problem 
would be the plane strain approach,  
corresponding to a long cylinder with circular cross-section.
To see how sensitive the compression experiment is to the geometry,
we compared this simple case 
(unrealistic for a cell) to
the axially symmetric case of 
a 3D sphere compressed between the plates. Note that the latter needs
solving all equations defined above in cylindrical coordinates.

Fig.~\ref{fig:figure5} A shows the distribution of stresses, visualised via $tr(\matr{\Sigma})$, 
within the cross-section 
of a cell in plane strain 
(top, cylinder geometry; note that this implies that the nucleus is also a cylinder)
and of a spherical cell 
in axial symmetry (bottom). 
The nuclear stiffnesses are $E_N/E_C=2$ (left) and $10$ (right), respectively. 
In the snapshots, the plates have a distance of 
$90\%$ of the initial cell diameter $2R_C$.
Both cases show an increased stress concentration for increasing nuclear stiffness in the regions between the nucleus and the plates, with a band-like stress accumulation connecting the cell edge in contact with the plates and the nucleus.
The plane strain case shows an overall higher stress, since it does not allow a considerable stress relaxation within the nucleus, leading to slightly higher cytoplasmic deformations and therefore a higher eccentricity of the cross-sectional shape as compared to the axially symmetric situation.
Nevertheless, overall the behavior is rather similar.

To further quantify the compression experiments, we obtained
the force-compression curves for the results shown in Fig.~\ref{fig:figure5}~A. 
This was done by calculating the total force $F=\int \lvert\nabla\cdot\Sigma\rvert\rho dV$ in mechanical equilibrium for the respective total compression $\delta$ of the cell, 
normalized by the 
cell diameter $2R_C$, as shown in Fig.~\ref{fig:figure5} B.
As can be noticed, a consistently higher force is required to deform a plane strain cylinder (triangles) by the same $\delta$ as compared to a sphere in axial symmetry (circles), consistent with Fig.~\ref{fig:figure5} A.
Note that for the resulting line contact problem in plane strain, 
the fundamental measure for this case is the in-plane force per length $F/L$.~\cite{Johnson_Contact_Mechanics}
In order to compare the force-compression curves in both geometries, we determined the 
length of the cylinder $L=4/3 R_C$ in plane strain, such that the cylinder volume is equal to the sphere volume in axial symmetry, 
and multiplied the average force per length by $L$.

Importantly, 
for both contact problems
studied here, plane strain and axial symmetry,
there exists an analytical solution
for the force-compression relation 
in the absence of the nucleus.~\cite{Johnson_Contact_Mechanics}  
The force-compression relation of an elastic sphere compressed by two rigid plates is the Hertz problem 
with $F \propto \delta^{3/2}$ for an arbitrary pressure distribution.~\cite{Hertz1882,Johnson_Contact_Mechanics}
In plane strain, the relation 
is more complicated and can be given as 
$\delta \propto (F/L)\log(B/\sqrt{F/L})$, where B is a constant containing information about the cell size and its effective stiffness.~\cite{Johnson_Contact_Mechanics}
Fig.~\ref{fig:figure5} B shows,
apart from the numerically obtained data (symbols), also fits to these relations (solid curves), resulting in a very good agreement for both geometries. 
Importantly, the Hertzian theory 
$F \propto \delta^{3/2}$ is still valid, even in the presence of a rather large and stiff nucleus.

As the two-plate setup is extensively used to 
measure cellular stiffnesses, 
we tried to infer the effective Young's modulus $E_{eff}$ 
(\textit{i.e.}~cell plus nucleus as measured in the respective apparatus) 
of our model cell in the physically relevant axial symmetric situation. 
We used the full Hertzian law $F=(\sqrt{2R_C}E'/3)\delta^{3/2}$ for a parabolic pressure distribution with $F$ the total force per plate and $E'=E_{eff}/(1-\nu_{eff}^2)$.
Here $E_{eff}$ and $\nu_{eff}$ 
are the effective elastic parameters of the cell-nucleus composite for rigid plates.~\cite{Hertz1882}
We assumed here that $\nu_{eff}=\nu_C=\nu_N$. 
Fig.~\ref{fig:figure5} C then shows that $E_{eff}$ increases non-linearly with increasing nucleus stiffness $E_N$.
For physiological nucleus sizes $V_N/V_C=0.125-0.3$, the effective modulus $E_{eff}$ experiences 
an up to three-fold increase for $E_N/E_C=10$.
Note, that for $E_N/E_C=1$ the comparison with Hertzian theory 
yields an effective modulus slightly $E_{eff}<1$, resulting from the unknown pressure distribution in the phase field simulation. 

As long as the plates are not strongly adhesive, 
compression should not activate contractility and thus 
above we did not include any active stress. If we include active
stress in our simulations, we find similar stress patterns, but the fit to 
Hertz-theory did not work well anymore (results not shown).

Our simulations can also be used to study
the effect of different cell shapes. Fig.~\ref{fig:figure5} D shows the compression of an initially pancake-like shaped 
cell in axial symmetry, similarly as studied in Ref.~\cite{Fischer-Friedrich_2016_rheology_of_active_cell_cortex_in_mitosis}, 
for different nucleus stiffnesses $E_N/E_C=1,2,10$. 
It can be directly compared to the initially spherical cell in Fig.~\ref{fig:figure5} A, bottom.
Again, for increasing nucleus stiffness a redistribution of stresses within the cell is visible. 
For $E_N/E_C=1$, the regions of highest stress are located close to the cell boundary at the transition points from vanishing to finite curvature, as predicted before.~\cite{Fischer-Friedrich_2016_rheology_of_active_cell_cortex_in_mitosis} 
However, for increasing nucleus stiffness the upper and lower poles of the nucleus, nearest to the plates, become the zones of  highest stress. Again a band-like stress from the cell boundary in contact with the plates to the nucleus boundary is visible. 
The main impact of cell morphology 
(sphere vs.~pancake) 
on nuclear straining thereby comes from the distance between the nucleus and the cell boundary:
forces are better propagated to the nucleus for flatter cell shapes.

\begin{figure}
    \centering
    \includegraphics[width=\linewidth]{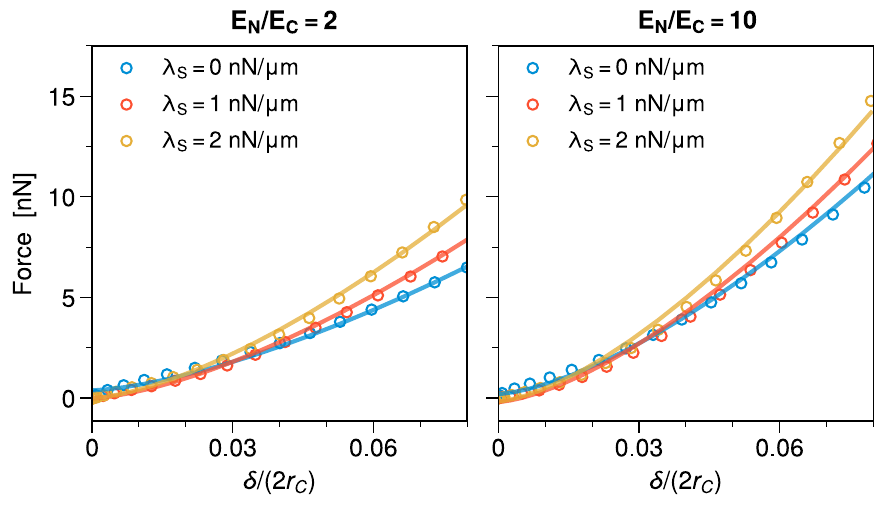}
    \caption{Force-compression curves for a spherical cell subject to a cortical surface tension of varying strength $\lambda_S=0,1,2\, {\rm nN/\mu m}$ for different nuclear stiffnesses $E_N/E_C=2$ and $10$.
    The symbols are the numerical results and the solid lines fits 
    according to the Hertz law.
    Note that the deformation of the  cell due to the cortical tension,
    prior to compression, 
    changes its diameter (from $2 R_C$ to 
    $2 r_C$), which 
    is used to normalize $\delta$. 
    The compression is quasi-stationary;
    parameters are $E_C=0.5\, {\rm kPa}$ and $\xi=0.04\, {\rm nN\cdot s/ \mu m^3}$ for better numerical stability.
    Other parameters as in Fig.~\ref{fig:figure5} and Table~\ref{tbl:parameters}.}
    \label{fig:surface tension}
\end{figure}

Finally, we studied the effect of
cortical tension on 
the force-compression curves 
of spherical cells.
Cortical tension is due to
myosin II motor activity in 
the actin cortex located directly
underneath the plasma membrane.
Since the cortex is thin compared 
to the cell dimension,
effectively this effect can be described  
as a surface tension $\lambda_S$.
Hence in the phase field 
sense \cite{Folch99,Benjamin_st}, 
we add the force
\begin{equation}
    \vec{F}_{st}=\lambda_S\kappa_\rho\frac{\nabla h(\rho)}{f(h(\rho))}
\end{equation}
to $\vec{F}_{tot}$ 
in Eqs.~\eqref{phase field eq. for rho},\eqref{elastodynamic eq.}. 

In the simulation,
the cell is first allowed  
to mechanically relax 
under the applied cortical tension, 
then the compression is started.  
Hence 
we now normalized
the compression height $\delta$ 
by the cell diameter $2 r_C$ 
in mechanical equilibrium, 
with applied surface tension but 
before compression.
We chose a lower cell stiffness than before,
$E_C=0.5\, {\rm kPa}$, to make the effect
more apparent, and a 
realistic cortical surface tension 
range of up to 
$\lambda_S=2\,{\rm nN/\mu m}$ 
\cite{cortex1,cortex2}.
Fig.~\ref{fig:surface tension} shows 
the obtained force-compression curves
for nuclear stiffnesses $E_N/E_C=2$ and $10$.
One can see 
that increasing $\lambda_S$ results in an
increase of the required force for compression,
especially for larger compression.
However, the stiffer the nucleus, 
the smaller is the effect relative 
to the case without surface tension.

To quantify, 
we again performed a comparison to Hertz 
theory (solid lines 
in Fig.~\ref{fig:surface tension}),
yielding good fits for all tested cases.
For the extracted effective cell
stiffnesses $E_{eff}$
we found an increase 
of $\sim 63 \%$
for $E_N/E_C=2$ 
(from $E_{eff}=0.53 
{\rm kPa}$
for $\lambda_S=0\, {\rm nN/\mu m}$ 
to $E_{eff}=0.82 
{\rm kPa}$ 
for $\lambda_S=2\, {\rm nN/\mu m}$)
and of $\sim25 \%$
for $E_N/E_C=10$ 
(from $E_{eff}=0.91 
{\rm kPa}$
for $\lambda_S=0\, {\rm nN/\mu m}$
to $E_{eff}=1.22 
{\rm kPa}$
for $\lambda_S=2\, {\rm nN/\mu m}$).
Hence cortical surface tension effectively 
stiffens cells, as is to be expected 
since compression increases the surface area.
The effect decreases with increasing
nuclear rigidity, and also with 
increasing cytoplasmic stiffness.

\subsection{Compressing cells: viscoelastic effects}
\label{comp_VE}

\begin{figure}[t!]
    \centering
    \includegraphics[width=\linewidth]{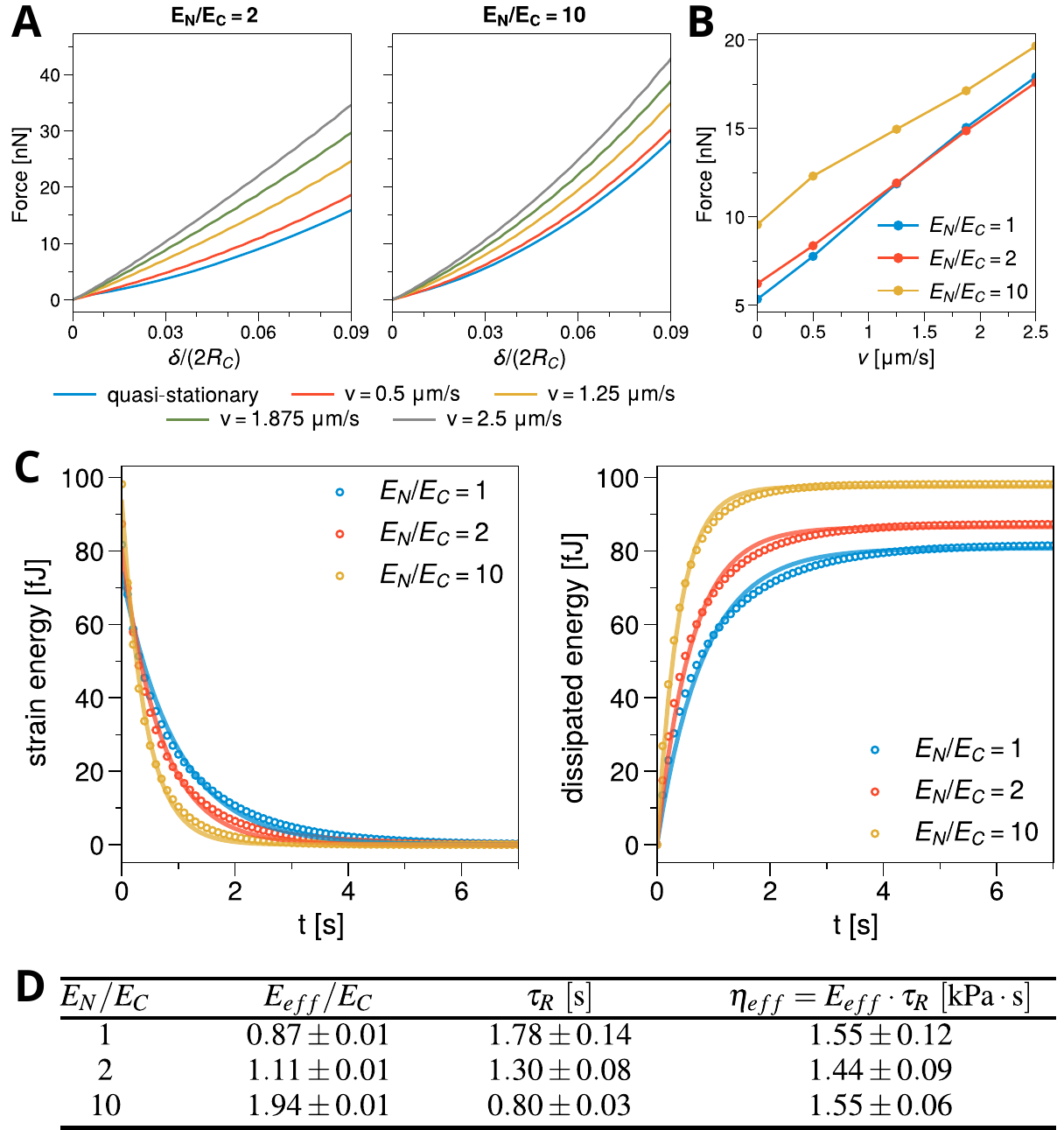}
    \caption{
    (A) Numerically obtained force-compression curves
    for a spherical cell between two plates in axial
    symmetry. The plates are moved 
    with a velocity 
    of $v=0, 0.5, 1.25, 1.875, 2.5\, \mu{\rm m/s}$ 
    (blue to grey); $v=0$ corresponds to 
    the quasi-stationary case studied in Fig.~\ref{fig:figure5}.
    Left panel: $E_N/E_C=2$; 
    right panel $E_N/E_C=10$.
    (B) Force-velocity curve extracted 
    from (A) at a given $\delta$, 
    here the last time point where 
    $\delta/(2R_C)\leq 0.05$. 
    The force required for the same
    compression increases with plate 
    velocity, as time becomes
    increasingly insufficient to relax stress. 
    (C) Shown is the strain energy 
    (left) and the dissipated energy (right),
    as a function of time 
    for different nucear stiffnesses 
    $E_N/E_C=1,2,10$. At $t=0$ 
    the plates were suddenly removed and
    the cell allowed to relax back 
    to its spherical, undeformed state. 
    The numerically obtained solutions
    (circles) are fit (solid lines) 
    allowing to extract
    relaxation times and effective viscosity.
    (D) Table giving $E_{eff}$ obtained from
    Fig.~\ref{fig:figure5}~D, 
    the average $\tau_R$ 
    from the fits in (C) and the resulting
    $\eta_{eff}=E_{eff}\cdot \tau_R$. 
    Parameters as in Fig.~\ref{fig:figure5} 
    with $V_N/V_C=0.125$, 
    i.e.~$R_N=6.25\, \mu{\rm m}$. 
    Other parameters as in table~\ref{tbl:parameters}.}
    \label{fig:figure_5_compr_ve}
\end{figure}

As shown in the previous section, 
the quasi-stationary (elastic) case 
is well suited 
to determine the effective stiffness of the 
cell-nucleus composite.
Rapid compression (\textit{i.e} faster 
than the elastic relaxation)
leads to viscoelastic effects.
We consider again
the case of a spherical cell with nucleus.
Now the plates are moved towards each other
in a continuous fashion with a velocity $v$.

Fig.~\ref{fig:figure_5_compr_ve}~A 
shows resulting force-compression curves 
for nuclear stiffnesses 
$E_N/E_C=2$ (left)  and $E_N/E_C=10$ (right) 
and for different plate velocities $v$.
The blue curve marks the quasi-static case
discussed in Fig.~\ref{fig:figure5}. 
One clearly sees that increasing the 
plate velocity
increases the force required to reach the same 
relative compression $\delta$.
We also note that if the plate motion 
is stopped in between, 
the force relaxes and reaches the 
corresponding lower bound given by 
the quasi-stationary case (blue curve),
as observed in experiments~\cite{Fischer-Friedrich_2016_rheology_of_active_cell_cortex_in_mitosis}.

In accordance to the previous results,
comparing the cases 
$E_N/E_C=2$ and $E_N/E_C=10$ of 
Fig.~\ref{fig:figure_5_compr_ve}~A 
shows that increasing the nuclear stiffness 
increases the force. 
On the other hand, the stiffer the nucleus, 
the smaller the effect of the 
compression velocity becomes.
To further quantify these observations, 
Fig.~\ref{fig:figure_5_compr_ve}~B shows
the force-velocity relation, obtained  
by calculating the total force experienced 
by the cell at a certain $\delta$ 
(here at the last time point where 
$\delta/(2 R_C)\leq 0.05$), 
for the plate velocities $v$ shown in
Fig.~\ref{fig:figure_5_compr_ve}~A.
An almost linear increase of the force 
with plate velocity $v$ is visible, 
with only a marginal difference 
between $E_N/E_C=1$ and $E_N/E_C=2$,
suggesting that soft nuclei have 
only a small effect.
The increase of the force with plate 
velocity $v$ is due to insufficient time 
of viscoelastic stress relaxation
to mechanical equilibrium
for finite plate velocity.
Such a higher (not completely relaxed) 
stress should consequently lead to the
higher effective stiffness for the 
cell-nucleus composite
at a given velocity. 
However, we refrained from fitting the 
force-compression curves to the Hertz law, 
as it is only valid in the stationary case.

As a second test to show that the 
phase field model describes viscoelastic
effects of Kelvin-Voigt-type correctly,
we analyzed a relaxation experiment. 
A spherical cell was compressed 
in a quasi-stationary manner as described 
in the last section. 
After reaching a compression 
$\delta$ of $10\%$ of the initial cell
diameter $(2R_C)$, 
the plates were removed instantly 
and the cell allowed to relax back 
to its initial, undeformed configuration. 
Fig.~\ref{fig:figure_5_compr_ve}~C 
shows the time evolution of the 
strain energy 
$E_{\epsilon}=(1/2)\int_V\Sigma_{ij}\epsilon_{ij} dV$
(left) and of the dissipated energy 
$E_d=E_0-E_{\epsilon}$ (right) 
from the time point 
of plate removal ($t=0$) of elastic 
energy $E_0=E_{\epsilon}(t=0)$, 
for different nuclear stiffnesses $E_N/E_C$.
The circles are the numerically obtained
results, showing that the strain energy
decays exponentially and that
the dissipated energy levels to a plateau 
for large times, both
indicating the mechanical equilibrium of 
the undeformed state.

As discussed already in Sect.~\ref{failadh},
in the viscoelastic regime
our phase field model is of Kelvin-Voigt type.
We hence can fit the corresponding 
strain energy function,
$E_{\epsilon}(t) = E_0 \exp(-2t/\tau_R)$, 
and dissipated energy function 
$E_{D}=E_0\left[1-\exp(-2t/\tau_R)\right]$,
to the numerical data, 
cf.~the solid curves in
Fig.~\ref{fig:figure_5_compr_ve}~C,
allowing to extract the relaxation 
timescale $\tau_R$ (here we average over 
the results from strain energy and 
dissipated energy) for the different 
nuclear stiffnesses considered.
Together with the effective 
cell-nucleus rigidities, 
obtained previously 
in Fig.~\ref{fig:figure5}~D 
for $V_N/V_C=0.125$, 
we are able to infer 
the effective viscosities 
$\eta_{eff}=E_{eff} \cdot \tau_R$.
The obtained values are given in the table
Fig.~\ref{fig:figure_5_compr_ve}~D.
As expected, the relaxation times decrease 
with nuclear stiffness but the effective 
viscosity $\eta_{eff}$ remains 
approximately constant, as it is
determined by $\xi$ 
in Eq.~\eqref{elastodynamic eq.}. 
This shows that $\xi$ sets the 
effective viscosity 
of the composite model under compression, 
similar as had been shown in
Sect.~\ref{failadh} for the adhesion geometry.
We note that we used a small $\xi$ here,
to get a relaxation on a time scale
of seconds, since the numerical time step 
in the  compression geometry is very small
($10^{-4}\rm{s}$). This value corresponds 
more to intracellular relaxation time scales, 
while those for a whole cell are 
of the order of tens of seconds or minutes
\cite{FabrycellVE}.
We stress that there is no 
problem to increase $\xi$, and consequently
the effective viscosity and $\tau_R$,
as exemplified in Fig.~\ref{fig:figure4}~B
for the adhesion geometry.

\begin{figure*}
    \centering  \includegraphics[width=\linewidth]{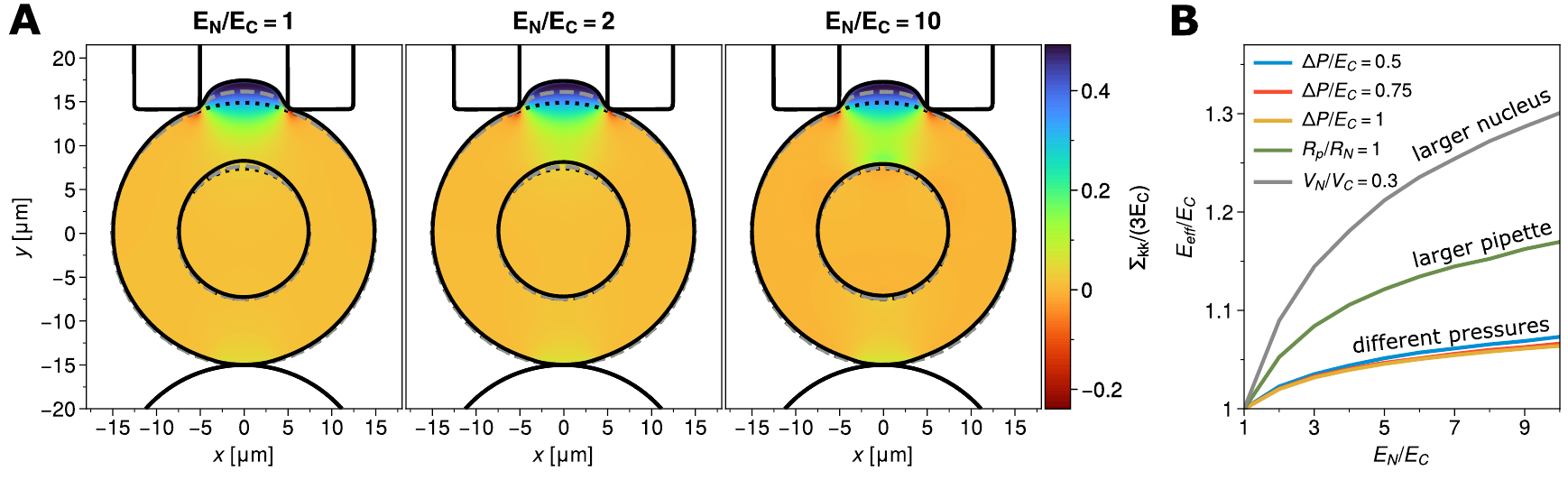}
    \caption{(A) Micropipette aspiration of a spherical cell ($R_C=15\, {\rm \mu m}$) with nucleus ($R_N=7.5\, {\rm \mu m}$) into a pipette of $R_p/R_N=2/3$ with $\Delta P/E_C=0.5$ in axial symmetry. Shown are the cases $E_N/E_C=1$, $E_N/E_C=2$ and $E_N/E_C=10$. Cell and nucleus boundary are depicted for three different time points: before pressure application (dotted black), during pressure application (dashed gray) and in mechanical equilibrium (solid black). The solid line outside the cell marks the edges of the pipette walls (top) and the spherical bead the cell adheres to (bottom).  The color map shows the trace of the stress tensor normalized by the cytoplasmic stiffness $E_C$. (B) Effective moduli extracted from experiments as shown in A
    at $\Delta P/E_C=0.5$ (blue),
    for higher pressures $\Delta P/E_C=0.75$ (red) and $\Delta P/E_C=1$ (yellow), 
    and at $\Delta P/E_C=0.5$
    but for a larger nucleus $V_N/V_C=0.3$ (gray) 
    or for a larger pipette $R_p/R_N=1$ (green).
    All cases show an increase in effective modulus with nucleus stiffness, but much smaller as  compared to the compression experiment in Fig.~\ref{fig:figure5} C. 
    Note, due to the unknown shape factor for the micropipette, we shifted the curves slightly such that for $E_N/E_C=1$ the expected modulus is recovered.
    Simulations were performed on $N=512\times512$ grid points. If not mentioned otherwise, the parameters used are $R_C=15\, \mu{\rm m}$, $R_N/R_C=0.5$, $E_C=1\, {\rm kPa}$, $\nu_C=\nu_N=0.48$, $\alpha=6\, {\rm kPa}$ and $Y=5\, {\rm nN/\mu m^2}$ (unit for adhesion strength is now different due to
    phase field-type definition, Eq.~\eqref{eq: pipette adhesion}). Other parameters as in table~\ref{tbl:parameters}.}
    \label{fig:figure6}
\end{figure*}

\subsection{Micropipette aspiration}
An alternative to cell compression experiments for measuring cellular rheological responses
are micropipette experiments.~\cite{Hochmuth_2000_micropipette_review,Evans_Yeung_1989,Herant_Dembo_2005_micropipette,Tinevez_Paluch_2009_micropipette,Robinson_2013_micropipette}
In this setup, cells are sucked into a pipette tube by applying a pressure difference $\Delta P$ between the tube's interior and the exterior space.
In this setup, forces are more locally applied compared to global straining  in compression experiments.
Micropipette aspiration has already been studied numerically and together with experiments 
showed that cells can
have elastic and viscous signatures.~\cite{Dembo_1999_micropipette} 
Therefore, this experiment has been  used to measure both the elastic modulus $E$ and viscosity $\eta$ of cells.~\cite{Evans_Yeung_1989}

We are again interested in the influence of the nucleus on the measurement of the effective cell stiffness, in the context of this more local force application.
Considering a spherical cell with axial symmetry, the stationary pipette wall can be modeled as in the previous example by using a field $\varphi(\vec{x})$, placing it closely to the cell membrane. 
Before sucking the cell into the pipette, we first let the cell relax into mechanical equilibrium due to the interaction with the pipette walls, \textit{cf.}~Eq.~\eqref{volume exclusion force}. To prevent any rigid body motion of the cell, we let it adhere 
to a sphere (field $\Tilde{\varphi}(\vec{x})$)
on the side opposite to the pipette, as also done experimentally~\cite{Chesla_1998_micropipette,Evans_2004_micropipette} 
using an adhesion force 
\begin{equation}\label{eq: pipette adhesion}
    \vec{F}_{adh}=Y(\nabla\Tilde{\varphi})(\nabla h(\rho))\vec{u}.
\end{equation}
Note that this is the phase field version of Eq.~\eqref{adhesion force},
modeling adhesion of the cell with strength $Y$ when it is in contact with the sphere.

Applying now a pressure, $P_1$, in the micropipette tube that is smaller than the pressure $P_0$ in the cell's interior (the outside pressure is assumed to be $P_0$ as well) 
leads to a boundary force acting at the cell membrane within the pipette like
\begin{equation}
    \vec{F}_p = \Delta P(\vec{x})\frac{\nabla h(\rho)}{f(h(\rho))}
\end{equation}
where $\Delta P(\vec{x})=(P_1-P_0)p(\vec{x})$ and where $p(\vec{x})$ marks the micropipette interior
(where $P_1$ is applied).

Fig.~\ref{fig:figure6} A shows results for the aspiration 
of spherical nucleated cells
($R_C=15\,{\rm \mu m}$, $R_N=7.5\,{\rm \mu m}$)
for nucleus stiffnesses $E_N/E_C=1,2$ and $10$ into a micropipette of radius $R_p=2/3 R_N=5\,{\rm \mu m}$ using a pressure difference of $\Delta P/E_C=0.5$. 
As expected, the highest positive stresses occur at the cell membrane within the pipette, while negative stresses arise at the edges of the micropipette, where it effectively pushes against the cell. 
Furthermore, in the case of stiff nuclei ($E_N/E_C=10$), stress accumulation occurs again in the vicinity of the nucleus boundary nearest to the pipette. 
This again suggests the possible perception of mechanical stimuli 
by the nucleus, even for very locally applied forces. 
While stiffer nuclei only deform marginally and are shifted 
within the cell towards the pipette position, soft nuclei ($E_N/E_C=1,2)$  show some egg-like asymmetry in their morphology due to deformation. 
The black curves in the panels of Fig.~\ref{fig:figure6} A show 
the cell and nucleus boundaries 
($0.5$-phase field isocurves) 
for three different time points, 
to exemplify the dynamic nature 
of the problem.

Also in the micropipette geometry,
one can extract an effective modulus for the cell-nucleus composite.
Within the elastic regime of aspiration, the stiffness can be approximated by the relation $E=(3\zeta/2\pi)\Delta P (R_p/L_p)$, where $R_p$ is the inner micropipette radius and $\zeta$ is a shape factor for the micropipette geometry.~\cite{Theret_1989_micropipette} No closed form exists for calculating the shape factor $\zeta$.

To calculate the effective cell stiffness $E_{eff}$ from the numerics, we determine the aspiration length $L_p$ and, knowing the applied pressure difference and the micropipette radius, we estimated the effective modulus over a range of nucleus stiffnesses, nucleus sizes, 
pressure differences 
and micropipette radii, \textit{cf.} Fig.~\ref{fig:figure6}~B.
All tested cases yield
$E_{eff}\approx1$ for $E_N/E_C=1$ with a deviation of less than $5\%$ for $\Delta P/E_C=0.5$.
Since the shape factor is unknown, we hence shifted all results such that for $E_C=E_N$ we get the correct modulus.
There also is a slight dependence 
on the applied pressure.
However, with increasing pressure (from blue to yellow curves in Fig.~\ref{fig:figure6}~B), the extracted effective moduli $E_{eff}$ approach each other, indicating that the method is best suited for 
sufficiently large applied pressures.

All studied cases show an increase in effective stiffness for stiffer nuclei. However, this increase is approximately $30 \%$ for the largest and stiffest nuclei.
Hence the localized force application
due to the micropipette geometry leads to much lower measured $E_{eff}$ 
compared to the global cell compression geometry, \textit{cf.}~Fig.~\ref{fig:figure5}. 
This clearly demonstrates -- and quantifies -- that experimentally measured effective cell stiffnesses 
do not only depend on the inner structure of the cell, but also 
on the experimental setup.

\section{Discussion and conclusions}

Cell mechanics and mechanotransduction are strongly influenced by the largest cellular organelle, the nucleus. 
Despite increasing evidence of its mechanical importance, models explicitly accounting for nuclear mechanics are still rare.
We here developed a two-phase field approach for modelling cell mechanics with an additional internal compartment associated with nuclear elastic properties and investigated the mechanical response of cells in a selection of biologically relevant geometries and experimental setups. We also verified our approach in several cases for which analytical solutions are available to the elastic equations and investigated the effects of cortical tension and viscoelasticity.

In the first part described in section \ref{Adhcell_const}, 
we considered the case of spread cells in unstructured and structured environments using a 2D plane stress formulation.
For homogeneous adhesion, the effect of a physiologically sized nucleus on the mechanical response of the cell is small as the adhesion to the substrate effectively shields the nucleus from deformations and stresses. Even peripheral adhesion on a ring pattern is still sufficient to protect the nucleus.
However, in more structured environments with highly localized adhesion sites, a much higher transmission of stresses to the nucleus was observed. 
These observations demonstrate theroretically
that the actual adhesion geometry can be sensed by 
cells at the nucleus, similar to recent conclusions with a purely elastic model (no phase field).~\cite{Solowiej-Wedderburn_Dunlop_2022_cell_adhesion_pattern}

When modeling micro-patterned environments, stiffer nuclei also change the cellular morphology, by perturbing the formation of the invaginated arcs.
Additionally, the nuclear position largely effects the stress distribution within the cell, which may be an important input for the cell with regard to the determination of its polarity, \textit{e.g.} when having to distinguish between front and back. 
For stiff nuclei, "stress bridges" resembling stress fibers form from close-by focal adhesions to the nuclear boundary, suggesting an effect on the perception of mechanical cues. Again a similar effect can be seen in purely
elastic models (no phase fields).~\cite{Ronan_2014} 

The here-proposed phase field method allows to model not only stationary but also dynamic situations. 
As a simple example we considered the failure of a focal adhesion for a cell on a hexagonal micro-patterned substrate.
The coupling of phase field dynamics and elasticity made it necessary to use an elastodynamic formulation for the evolution of the displacement field.~\cite{Chojowsi_2020_EPJE} 
We here showed, that the relaxation into mechanical equilibrium is of Kelvin-Voigt type.
In turn, if a purely elastic behaviour of the system is desired, it should not be probed on timescales shorter than the respective relaxation time.
It should also be noted that the elastic description memorizes 
the initial condition (i.e.~the reference state of the elastic displacement) of the cell before the 
application of forces or stresses.
Hence in situations such as the study of adhesion
failure in Fig.~\ref{fig:figure4} A,
the cell does not relax to a (deformed) pentagonal shape with an invaginated arc at the cell edge where the disappeared focal adhesion was located. In the future, the memory effect can be removed by
an extra dynamics for the reference state.

In the second part, described in section
\ref{confined},
we modeled compression experiments of cells between two parallel plates and the aspiration of cells into micropipettes in an axial symmetric geometry.
Again, stiffer nuclei showed stress accumulation near their boundary pointing towards a significant role of nuclear mechanics in determining the properties of the cellular environment.
Importantly, our model allowed the extraction of effective elastic moduli of the cell-nucleus composite for both experimental methods, yielding consistently lower effective moduli for local pressure application in micropipette experiments compared to more global cell compression. 
This shows theoretically that the determination of effective cell moduli is not only dependent on cell geometry but also the experimental setup used. A similar conclusion has been reached when experimentally comparing
different methods to probe whole cell mechanics.~\cite{wu2018comparison}
For the micropipette aspiration experiments 
the extraction of elastic moduli is best suited for sufficiently large pressures. Low pressure application leads to a slight underestimation in the range of $5$\% of the effective cell stiffness, which is partially influenced by the unknown shape factor for the pipette, cf.~the
discussion of Fig.~\ref{fig:figure6} B.
We also investigated the effect
of cortical tension in the compression geometry. 
It results in an increase of the required 
force needed for compression. 
The stiffer the nucleus, 
the smaller is the effect relative 
to the case without cortical tension.

The quasi-stationary
compression of cells is described well by Hertz theory
and therefore allows the identification of an effective 
modulus, similar to a very recent computational study with
elasticity (no phase field).~\cite{wohlrab2024mechanical} 
We next demonstrated that our phase field approach 
is also applicable in the viscoelastic regime, e.g.\ for dynamic compression with different plate 
velocities and relaxation studies.
The force required for the same compression 
increases with plate velocity, as time
becomes increasingly insufficient to relax 
the stress induced by the plate motion. 
From numerical relaxation experiments 
we could extract the relaxation time scale,
which can be adjusted in the model varying the
parameter $\xi$ in 
Eqs.~(\ref{phase field eq. for rho}), 
(\ref{elastodynamic eq.}), 
and the effective viscosity, confirming that
our approach is fully consistent with the
Kelvin-Voigt viscoelastic solid. At the current
stage, our model does not describe 
viscoelasticity of Maxwell type, 
that is a viscoelastic fluid
without memory. The best way to achieve this 
in our context
might be to introduce an own dynamics for the reference state.

In the future, the here-developed method should prove useful for investigating the effect of a nucleus and/or other cellular organelles, potentially described with different material laws, in a large variety of situations. 
Additional new insights on mechanotransduction could be gained by examining the effect of the nuclear position within the cell in fully three-dimensional (3D) situations.
In this respect, the phase field method can be extended relatively easily to 3D, making it possible to consider more complex environments like fibrous network geometries 
or non-symmetric constrictions.~\cite{Moure_Gomez_2017,Benjamin3D_2019,
3Dprep}
The presented method could also be used to describe the role of cell nucleus mechanics in tissues, 
using the multi-phase field approach.~\cite{Nonomura_multi_PF_2012,LoeberSR_2015,Wenzel_2021_multi_PF_cell_migration} 
Note that recently, a new jamming transition due to the presence of nuclei was predicted for tissues 
by an active foam model.~\cite{Campas_2022_arXiv} It would be interesting to study
the same effect in our dynamic continuum framework. 

Another important context of cell and nuclear mechanics is cell migration through constrictions,
where the minimal constriction size is predominantly determined by the nuclear size and stiffness.~\cite{Thiam_2016_Perinuc_Arp2_3_actin_polym_enables_nucl_deformation_to_faciitates_cell_migration}
We envision to supplement the current approach by self-organized internal driving forces inducing cellular motility,  
that could be implemented 
by an actin "polarization" field~\cite{Ziebert2012} and should naturally 
enter the elastodynamic equation \eqref{elastodynamic eq.}.
In the context of mechanotransduction, the coupling of the proposed method to a system of reaction-diffusion equations should allow to model nuclear translocation of proteins like YAP/TAZ in response to nuclear straining (and opening of nuclear pore complexes~\cite{Elosegui_Artola_2017_force_triggers_YAP_nuclear_entry_across_NPC,Andreu_Roca_Cusachs_2022_mechanical_force_to_nucleus_regulates_nucleocyto_transport, Zimmerli_2021_NPC_dilate_and_constrict}). This could elucidate further 
-- and more directly --
the role of nuclear mechanics on spatio-temporal import dynamics and mechanically induced signalling events. 

In summary, the elastic phase field approach for modelling the mechanics of nucleated cells 
is very versatile and easy to generalize for future applications.
The results presented should be useful to quantify experiments 
and last but not least point to many interesting implications with regard to the role of the nucleus on whole cell mechanics, mechanosensing and related subjects. 

\section*{Appendix}

\subsection*{A \quad Other parameters of the model}
Table \ref{tbl:parameters} contains
the default values for additional parameters used, if not specified otherwise in the figure captions.

\begin{table}[h]
\small
  \caption{  \ Phase field parameters not specified in the figure captions}
  \label{tbl:parameters}
  \begin{tabular*}{0.48\textwidth}{@{\extracolsep{\fill}}llll}
    \hline
    Parameter & Symbol & Value & Unit\\
    \hline
    time step & $\Delta t$ & $0.001$ &${\rm s}$ \\
    diffusion coeff. phase field  (PF)$^{\ast}$ & $D_{\phi}$ & $1.25$ & ${\rm \mu m^2/s}$ \\
    friction coefficient$^{\ast\ast}$ & $\xi$ & $0.004$ &${\rm nN\cdot s/\mu m^3}$\\
    local suppression coeff.$^{\ast\ast}$ &$\gamma(\vec{x})$ & $0.014-0.04$ &${\rm nN\cdot s/\mu m^3}$\\
    regularization parameter & $\epsilon$ & $0.0025$ & ${\rm \mu m^2}$\\
    diffusion coeff. adhesion PF  & $D_Y$ & $0.25$ & ${\rm \mu m^2/s}$\\
    diffusion coeff. obstacle PF & $D_\varphi$ & $0.625$ &${\rm \mu m^2/s}$\\

    \hline
  \end{tabular*}
  
\vspace{3mm}

\noindent
$^{\ast}$For $\phi\in\{\rho,\psi\}$, i.e. cell and nucleus. Sets the interface width to $0.5{\rm \mu m}$.
\newline
$^{\ast\ast}$Note, in plane strain and axial symmetry the unit is $\, {\rm nN\cdot s/\mu m^4}$.
\newline In Sect.~\ref{confined} $\Delta t = 10^{-4} {\rm s}$.

\end{table}

\subsection*{B \quad Analytical solution for an adherent contractile disk-like cell with nucleus}
\label{appendix: ana. sol. }
We consider a concentric cell and its nucleus with possibly different Young's moduli $E_N$ and $E_C$ and Poisson's ratio $\nu_N$ and $\nu_C$ as depicted in Fig.~\ref{fig:figure2} A. Further we assume a homogeneous and isotropic contractile stress $\sigma_0(\vec{x},t)=\sigma_0$ and spring stiffness density $Y(\vec{x},t)=Y$. Both nucleus and cytoplasm are assumed to be linearly elastic. In each of these two cell compartments the equation of mechanical equilibrium
\begin{equation}\label{appendix: mech equi}
    \nabla \cdot \matr{\sigma} - Y \vec{u} = 0
\end{equation}
has to be solved under respective boundary conditions. From the radial symmetry of the problem follows that the only non-vanishing displacement is in the radial direction, \textit{i.e.~}$\vec{u}=u_r\vec{e}_r$. Therefore, Eq.~\eqref{appendix: mech equi} can be rewritten in polar coordinates yielding
\begin{equation}\label{diff eq. 2D cell+nucleus+substrate disp formulation}
	r^2\frac{\partial^2 u_r}{\partial r^2}+r\frac{\partial u_r}{\partial r}-\left(1+\frac{r^2}{l_\alpha^2}\right)u_r=0
\end{equation}
with $l_\alpha=\sqrt{E_\alpha d/\left[Y\left(1-\nu_\alpha^2\right)\right]}$ being the localization lengths~\cite{EdwardsSchwarz} for cytoplasm ($\alpha=C$) and nucleus ($\alpha=N$), respectively. Note, that $Y$ could also vary between cytoplasm and nucleus.
Eq.~\eqref{diff eq. 2D cell+nucleus+substrate disp formulation} is a modified Bessel equation with general solution
\begin{equation}\label{gen sol 2D cell+nucleus+substrate}
	u_r(r)=AI_1\left(\frac{r}{l_\alpha}\right)+BK_1\left(\frac{r}{l_\alpha}\right)
\end{equation}
where $I_1(x)$ and $K_1(x)$ are modified Bessel function of the first and second kind and $A$ and $B$ are constants to be determined via the boundary conditions for both subdomains. 
These are for the nucleus compartment
\begin{equation}
	u_r^N(0)=0 \qquad \text{and} \qquad \sigma_{rr}(R_N)=-\sigma_Nd ,\
\end{equation}
where $\sigma_N$ is the a priori unknown stress at the cell-nucleus interface,
and for the cytoplasm 
\begin{equation}
	 \sigma_{rr}(R_C)=-\sigma_0d \qquad \text{and} \qquad u_r^N(R_N)-u_r^C(R_N)=0 .\
\end{equation}
The stress $\sigma_N$ at the nucleus-cytoplasm boundary can be determined by using that 
\begin{equation}\label{stress at nucleus interface}
	\sigma_{rr}^N(r)\Big\rvert_{r\rightarrow R_N}=\sigma_{rr}^C(r)\Big\rvert_{r\rightarrow R_N} .\
\end{equation} 
By using the above boundary conditions, the resulting solution is 
\begin{equation}\label{full analytical solution}
	u_r(r)=
	\begin{cases}
		-\frac{\sigma_Ndl_N}{2\mu_N+\lambda_N}\frac{I_1\left(\frac{r}{l_N}\right)}{I_0\left(\frac{R_N}{l_N}\right)-\frac{2\mu_N}{2\mu_N+\lambda_N}\frac{l_N}{R_N}I_1\left(\frac{R_N}{l_N}\right)} \quad \text{for $0 \leq r \leq R_N$} 
		\\
		\\
		A_C I_1\left(\frac{r}{l_C}\right)
			+B_C K_1\left(\frac{r}{l_C}\right)  \quad \text{for $R_N < r \leq R_C$} 	
	\end{cases}
\end{equation}
with lengthy but straightforward to obtain expressions for $A_C$, $B_C$.

\section*{Author Contributions}
R.C., U.S.S.~and F.Z.~developed the model together. 
R.C. performed the calculations, developed the numerical code and analyzed the data. U.S.S.~and F.Z.
designed and supervised the research. All authors wrote the paper.

\section*{Conflicts of interest}
There are no conflicts to declare.

\section*{Acknowledgements}
This work was funded by the Deutsche Forschungsgemeinschaft (DFG, German Research Foundation) under Germany's Excellence Strategy EXC 2181/1 - 390900948 (the Heidelberg STRUCTURES Excellence Cluster). USS is a member of the Interdisciplinary Center for Scientific Computing (IWR).



\balance


\bibliography{nucleusSM.bib} 
\bibliographystyle{rsc} 

\end{document}